\documentclass[12pt]{article}

\usepackage{array}       
\usepackage{multirow}    
\usepackage{makecell}    
\usepackage{diagbox}     
\usepackage{float}

\usepackage{newtxtext,newtxmath}
\usepackage{caption}
\usepackage{graphicx}

\usepackage[letterpaper,margin=1in]{geometry}

\renewenvironment{abstract}
	{\quotation}
	{\endquotation}

\date{}

\makeatletter
\renewcommand{\fnum@figure}{\textbf{Figure \thefigure}}
\renewcommand{\fnum@table}{\textbf{Table \thetable}}
\makeatother

\usepackage{cite}

\usepackage{url}

\def\scititle{
	Observation of quantum free fall and the consistency with the equivalence principle
}
\title{\bfseries \boldmath \scititle}

\author{
  Or Dobkowski$^{1}$, Barak Trok$^{1}$, Peter Skakunenko$^{1}$, \and
  Yonathan Japha$^{1}$, David Groswasser$^{1}$, Maxim Efremov$^{2, 3}$,
  Chiara Marletto$^{4}$,\and Ivette Fuentes Guridi$^{5,6}$, 
  Roger Penrose$^{7}$, Vlatko Vedral$^{4}$,\and Wolfgang P. Schleich$^{3,8}$, Ron Folman$^{1}$\and \small $^{1}$ Ben-Gurion University of the Negev, Department of Physics \and \small and Ilse Katz Institute for Nanoscale Science and Technology, Be'er Sheva 84105, Israel\and
  \small$^{2}$ German Aerospace Center (DLR), Institute of Quantum Technologies, 89081 Ulm, Germany\and
  \small $^{3}$ Institut für Quantenphysik and Center for Integrated Quantum Science and Technology (\textit{IQST}), \and \small Universität Ulm, 89081  Ulm, Germany\and
  \small $^{4}$ Clarendon Laboratory, University of Oxford, Parks Road, Oxford \small OX1 3PU, United Kingdom\and
  \small $^{5}$ School of Physics and Astronomy, University of Southampton, \small Southampton SO17 1BJ, United Kingdom\and
  \small $^{6}$ Keble College, University of Oxford, Oxford OX1 3PG, United \small Kingdom\and
  \small $^{7}$ Mathematical Institute, Andrew Wiles Building, University of \small Oxford, Radcliffe Observatory Quarter, \and \small Woodstock Road, Oxford, OX2 6GG, United Kingdom\and
  \small $^{8}$ Hagler Institute for Advanced Study at Texas A\&M University, Texas A\&M AgriLife Research,\and \small Institute for Quantum Science and \small Engineering (IQSE), and Department of Physics and Astronomy,\and \small Texas A\&M University, College Station, Texas 77843-4242, USA\and
}

\begin{document} 

\maketitle

\pagebreak
\begin{abstract} \bfseries \boldmath
The unification of quantum theory and the general theory of relativity – describing gravity, is one of the most important challenges in science. Einstein’s general theory of relativity is based on the principle of equivalence, and has been confirmed to great accuracy for large bodies. However, in the quantum domain the equivalence principle has been predicted to take a unique form involving a gauge phase, equal to the quantum phase of a free-falling object. To measure this phase, we realize a novel cold-atom interferometer in which one wave-packet stays static in the laboratory frame while the other is in free fall. The observed relative-phase of the wave-packets confirms the predicted phase of a free-falling object, and shows that in our low energy regime, the equivalence principle may be applied to the quantum domain. Our observation constitutes a fundamental test of the interface between quantum theory and gravity. The new interferometer also opens the door for further probing of the latter interface, as well as to searches for new physics.
\end{abstract}

\noindent
\section{Introduction}
The two pillars of twentieth-century physics---quantum mechanics (QM) and general relativity (GR)---have long stood side by side, resisting all attempts to bring them convincingly together. Related to the latter, ever since Galileo and Newton it was clear that the phenomenon of free fall in a gravitational field is a paradigm of physics and must be fully understood as part of any description of nature. Einstein went a step further and used the geometry of space-time to describe gravity and free fall. 
An additional outcome of GR is that one should be able to describe nature from different frames of reference. Rooted in Galileo’s insight and elevated by Einstein to a foundational axiom, the principle of equivalence has guided our understanding of gravitation for over a century\,\cite{Will2012,Will2014,Nauenberg2016,Okon2011}. The rise of quantum mechanics has not changed any of these convictions. The connection between these two pillars of modern physics, remains one of the most important open questions in physics, and it is thus of paramount importance to understand the phenomenon of free fall and the equivalence principle (EP) also in the quantum domain.

The phase of free fall is predicted in a purely quantum manner to have a dependence $\frac{m}{6}g^2T^3/\hbar+gmzT$ on the free-fall time $T$, where $m$ is the mass of the object, $g$ is the gravitational acceleration relative to the surface of Earth, and $z$ in the spatial coordinate in the direction of gravity. This prediction follows the calculated phase accumulated by an object accelerating in a linear potential, and has been made starting from almost one hundred years ago by Darwin, Kennard and others\,\cite{Darwin1927,kennard1,kennard2,Greenberger1979,Beyer1986,Penrose1,Penrose2,Penrose3,Zimmermann2017}. Alternatively, one may derive this phase in a completely different manner, specifically, utilizing the EP and a Galilean transformation. The identity of the outcome of the two calculations may be seen as what allows QM and the EP to coexist. Indeed, numerous authors have suggested that this phase has implications regarding the EP in the quantum domain\,\cite{Greenberger1979,Nauenberg2016,Brukner2019,howl2019, marletto2020}. Surprisingly, this crucial phase has never been measured.

Matter-wave quantum interferometers are high-precision tools capable of measuring global and relative accelerations. Such interferometry in the presence of the gravitational field of Earth has been previously achieved with neutrons in the celebrated Colella-Overhauser-Werner (COW) experiment\,\cite{Colella1975,Rauch2015} (as well as by Bonse and Wroblewski for a noninertial frame\,\cite{Bonse1983}), and with atoms\,\cite{Chu0,Tino2021,Lellouch2022}, where the gravitational acceleration $g$\,\cite{Chu1,Lin2011,ChenYang2025,Cassens2025} and the gravitational constant $G$\,\cite{Kasevich2007,Muller2024} have been measured, but as noted, the explicit phase of a free-falling wave packet (WP) with respect to Earth has not been directly measured.  

Here we report the measurement of the quantum phase of an object in free fall (in the sense that the experimentalist does not apply any forces to it), as observed in a frame stationary with respect to Earth. A direct measurement of this phase requires a unique interferometer employing a superposition of one WP in free fall, and another WP stationary in Earth’s frame. We have named this interferometer the Quantum Galileo Interferometer, or QGI\,\cite{Galileo-footnote}.

Fig. 1(A) describes such a {\it Gedanken} (thought) experiment, where the WP that is at rest in the laboratory (Earth) frame is termed the “reference WP” while the WP that is free falling in this frame is termed the “ballistic WP”\,\cite{frames-footnote}.

\begin{figure}[H]\label{fig:Figure1V3.png}
\centering
\includegraphics[width=1\textwidth]{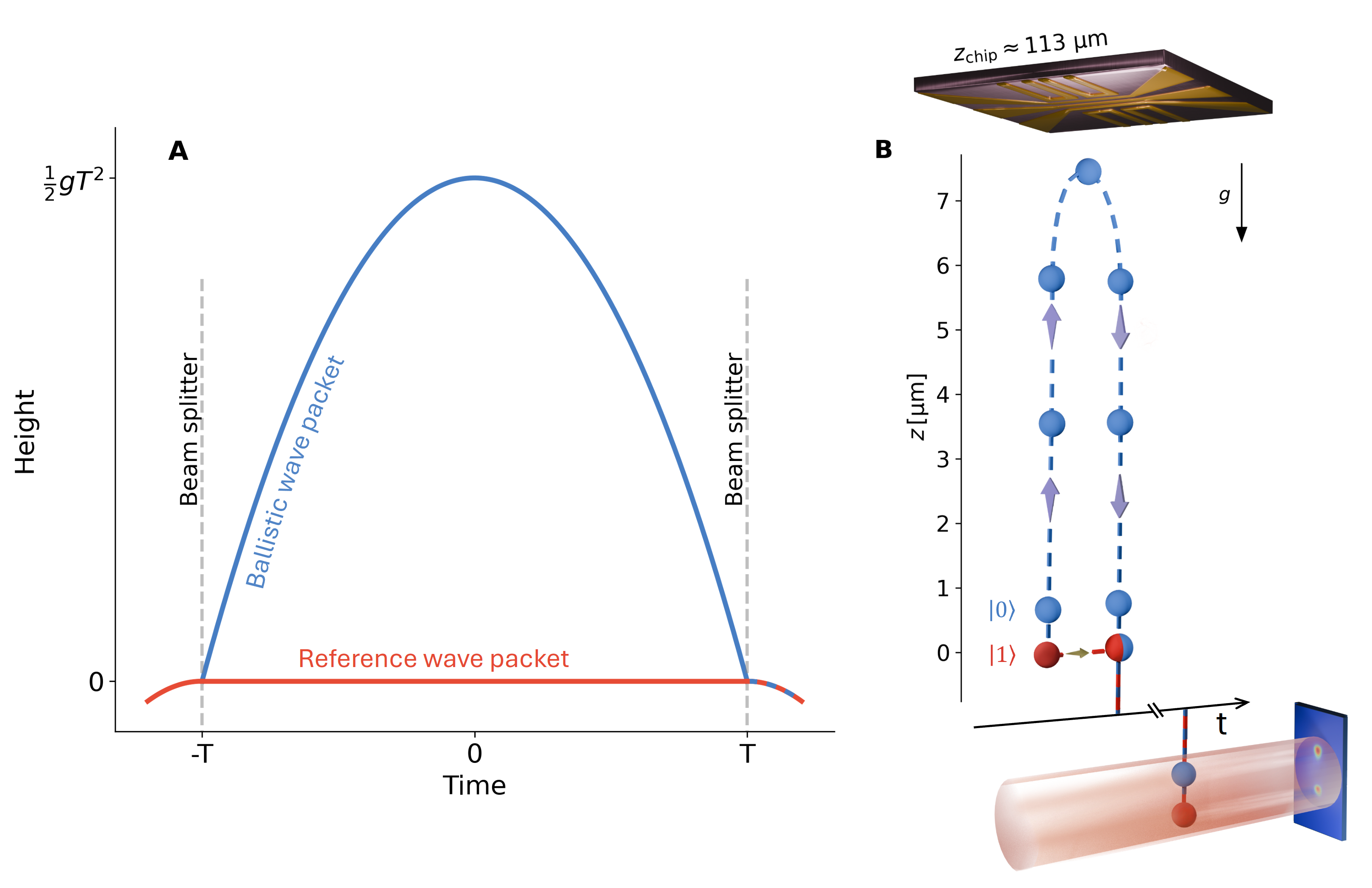}
\caption{
{\textbf{The QGI experiment to measure the phase accumulation of a free-falling particle.
}}
\newline
(\textbf{A}) Space-time diagram illustrating the classical trajectories of the WPs in the interferometer. A beam splitter creates a coherent superposition of two WPs with different spin states, launching one WP into a freely-falling ballistic trajectory for a duration of $2T$, while the second WP is held stationary, serving as a reference. A second beam splitter recombines the two WPs at the end of the ballistic trajectory. State-dependent detection (not shown) measures the population in each state, revealing the phase difference between the two trajectories. (\textbf{B}) A schematic of our experimental realization: In the longitudinal (1D) QGI interferometer, the stationary WP is held against gravity using magnetic gradients produced by currents in microfabricated wires on the atom chip. This WP, associated with state $|1\rangle \equiv|F=2,m_F=1\rangle$ of a $^{87}$Rb atom, is positioned $\approx 113\,\rm\mu m$ below the chip surface. The ballistic WP, associated with state $|0\rangle \equiv|F=1,m_F=0\rangle$, travels upwards reaching, in this example, a splitting of about $7.5\,\rm\mu m$ [about seven times the WP size (SM)]. Interchanging currents in the three central wires (shown in the figure) create a 2D quadrupole to realize a strong magnetic gradient with a small magnetic field, which minimizes phase noise. Finally, in order to measure the population in each spin state, another (reversed) magnetic gradient separates the two states in position so that they can be imaged by a CCD. See the Methods section for more details.}
\end{figure}


While a cubic phase dependence on the interferometer time may also appear in a few situations that are not related to free fall\,\cite{McDonald2014,dutta2016,Asenbaum2017,Rozenman2019,Amit2019}, observing this free-fall phase in the QGI configuration constitutes a direct test and confirmation of the phase of a free-falling object, and as noted, of the predicted phase when applying the EP to a quantum WP in free fall. Specifically, none of the previous experiments measuring a $T^3$ behavior, have had in the two interferometer arms the accelerations of $0$ and $g$ with respect to Earth, and consequently, none measured the predicted prefactor of $\frac{m}{6}g^2/\hbar$. As explained in the following, this requires a hybrid interferometer to be realized, a very different configuration from the latter interferometers.

To briefly show how the same phase is predicted by the EP, we will assume that the EP holds in the quantum domain, and calculate the observable consequence of such an assumption. To begin with, throughout this work we assume the equality of the inertial mass ($m_i$) and the gravitational mass ($m_g$), such that $m_i=m_g=m$, as suggested by Galileo and confirmed experimentally with early experiments by Newton, Bessel and E\"otv\"os and more recent experiments with improved accuracy\,\cite{Wagner2012,Touboul2017,Rosi2017,Asenbaum2020}. We of course also assume the universality of free fall to hold, so that also a $^{87}Rb$ atom in free fall accelerates towards Earth with acceleration $g$.

We now view the experiment from two frames of reference, one in free fall and one static relative to Earth. These two frames may be referred to as the Einsteinian and the Newtonian frames, respectively\,\cite{howl2019}. The transformation between the wavefunction $\psi_N$ in a Newtonian (laboratory) frame of reference, where the acceleration of a free object is denoted as $\ddot z=-g$, and the wavefunction $\psi_E$ of the same object in an Einsteinian frame that free-falls with the same acceleration $-g$, and when at $t=0$ both have the same velocity, involves a transformation phase which may also be referred to as a gauge phase $\phi_{\rm gauge}$\,\cite{marletto2020}. This gauge phase may be calculated in ways that do not rely on the EP, for example, by using Galilean transformations [see Supplementary Material (SM)]. 

If one then assumes that the inertial mass is equal to the gravitational mass, this transformation phase takes the form (e.g., \cite{howl2019,marletto2020}):
\begin{equation}
\psi_N(z,t)=e^{i\phi_{\rm gauge}(z,t)}\psi_E(z+g\frac{t^2}{2},t)\,,
{\rm where}\,\, \phi_{\rm gauge}=\frac{m}{\hbar}\left(-\frac16 g^2t^3-g z t\right)\,.\label{eq:phi_gauge} 
\end{equation}

According to the EP, $\psi_E$ is the wavefunction of a free particle that feels no forces in the free-falling (Einsteinian) frame\,\cite{Elevator_Remark}. Hence, if one assumes that the EP holds, this WP does not accumulate any phase (beyond free evolution). However, Eq.\,(1) implies that such a wavefunction accumulates a phase $\phi_{\rm gauge}$ in the laboratory (Newtonian) frame, simply due to the transformation between frames.

In conclusion, when comparing a quantum state in two different descriptions---one in which the gravitational field is treated as an external force, and another adopting the freely falling frame---a unique phase factor emerges, $m g^2 t^3 /6\hbar$\,\cite{Greenberger1979,Nauenberg2016,Brukner2019,howl2019,marletto2020}. This contribution, first identified nearly a century ago as the phase of a freely-falling object, has since attracted considerable theoretical scrutiny\,\cite{Darwin1927,kennard1,kennard2,Greenberger1979,Beyer1986,Penrose1,Penrose2,Penrose3,Zimmermann2017}. However, while this phase was typically calculated as the phase of an accelerating particle, without regard for the problem of the interplay between QM and GR, remarkably, as highlighted by the numerous works\,\cite{Greenberger1979,Nauenberg2016,Brukner2019,howl2019,marletto2020}, without this phase QM and the EP do not coexist. Specifically, by using the EP, as we did above, one predicts the same phase. In the case that the quantum calculation of an accelerating object under gravity, as was calculated by the numerous authors, was not complete, or in case the application of the EP to the quantum domain, as we have done, is lacking, it is of paramount importance to experimentally confirm this phase. We emphasize that in this work we will not be excluding all possible models for which the EP does not hold, as such models may also predict the same phase. Assuming the EP holds and observing no contradiction, we are able to show that the EP may be applied to a quantum WP.
Hence, beyond the measurement of the phase of a free-falling quantum object, the QGI enables the observation of the phase enabling the coexistence of QM and the EP. 

\section{Experiment}

Our experiment intended to measure this phase — by way of interference with a reference WP — follows the {\it Gedanken} experiment presented in Fig. 1(A) where at time $t = t_0$ the WP is split into two paths where in one path it is at rest in the Newtonian frame, due to an applied levitating force that is
exactly opposite to gravity (hereafter levitation condition), while in the other path it is in free fall, namely, at rest in the freely-falling Einsteinian frame. To achieve recombination of the two paths, the WP in the freely-falling path is initially launched upwards by a magnetic-gradient pulse with a 
velocity $v_0$ in the direction opposite to gravity and then falls freely for a time $2T = 2v_0/g$ (hereafter closing condition), after which an identical magnetic-gradient pulse erases the velocity difference between the two WPs so that the two paths overlap in position and momentum.

For an intuitive understanding of the QGI phase in the Newtonian frame, we fix the reference WP center at vertical position $z = 0$ so that it does not accumulate any propagation phase (beyond the free evolution of the WP), and calculate the phase accumulated by the ballistic WP. As noted, the latter accumulates no propagation phase in the freely-falling (Einsteinian) frame, so that the phase at the center of the ballistic WP in the Newtonian frame is nothing but the gauge phase. The phase of the ballistic path is the difference between the phase of the WP center at the two path endpoints, which is equal to the gauge phase at these points. Taking the middle of the path to be at time $t=0$, these endpoints are at $t=\pm T$, where the WP is centered at $z=0$ after a free-fall duration $T$ with respect to $t=0$ (Fig.\,1(A)). The gauge phase at the endpoints and the phase of the ballistic path then read
\begin{equation}
 \phi_{\rm gauge}(\pm T)=\mp \frac{m}{6\hbar} g^2T^3\, {\rm ;}\,\,
\phi_{\rm ballistic}=\phi_{\rm gauge}(T)-\phi_{\rm gauge}(-T)=-2\cdot\frac{m}{6\hbar}g^2T^3\,. \label{eq:ballistic}
\end{equation}
As $\phi_{\text{reference}}=0$, it follows that $\phi_{\text{ballistic}}$ is the relative phase observed in the experiment\,\cite{observable-footnote}. This phase is of course equal to the action of the ballistic path in the Newtonian frame, or equivalently of the reference path in the Einsteinian frame, as we show in the SM. In the SM, we also show that it can be calculated using the Galilean transformation. Note also that the linear term in Eq.\,(1) vanishes when the endpoints are at $z=0$  and even if the levitation condition is not satisfied it is cancelled by the magnetic gradient pulses, which are necessary for the closing condition (SM).  

Beyond the above description of the fundamental uniqueness of the QGI, in which the measured phase is equal to the free-fall phase, its technical distinctiveness should also be emphasized. While the Ramsey-Bordé interferometer is mostly characterized by the constant spatial separation between the two interferometer paths\,\cite{Borde}, and the Kasevich-Chu interferometer is mostly characterized by the piece-wise constant velocity separation (difference) between the two paths\,\cite{Chu2}, the QGI enables the control of all degrees of freedom, obtaining spatial, velocity and acceleration differences, almost at will. This allowed us to have one WP at rest and the other free falling. Furthermore, typical atom interferometers are free-space interferometers, whereby during most of their motion, the WPs do not experience any potential applied by the experimental setup\,\cite{Kovachy2015}. Some interferometers do have a potential applied to both WPs during their evolution, and these may be termed, guided interferometers\,\cite{Xu2019,Boshier2023}. In our case, both methods are used simultaneously, one for each of the WPs. The QGI is thus a hybrid interferometer.

In Fig.\,1(B) we present the experimental setting in which the QGI, utilizing $^{87}$Rb atoms, is conducted about $113\,\mu$m below an atom chip\,\cite{Keil2016}, where the current-carrying wires give rise to the magnetic field gradients which are responsible for the Stern-Gerlach (SG) type forces applied to the WPs. The QGI is a novel type of SG interferometer (SGI), leveraging a decade of SGI experiments in which the necessary techniques were developed\,\cite{keil2021}, and consequently, is now capable of demonstrating enhanced stability, larger phase accumulation, and especially the flexibility required to enable one WP to take a ballistic trajectory while the other is stationary.

During the free-fall time, the ballistic WP is in state $|0\rangle \equiv|F=1,m_F=0\rangle$, so that to first order in the Zeeman interaction it is not affected by the magnetic gradients. The reference WP is in state $|1\rangle \equiv|F=2,m_F=1\rangle$ and is held stationary by a ``holding magnetic gradient pulse" producing a magnetic acceleration $a$ in the opposite direction to gravity. Both states belong to the atomic ground state, whereby $F$ is the total angular momentum of an atomic state and $m_F$ is its projection on the quantization axis. The acceleration $a$ is independently tuned by an experimental procedure (see Methods and SM) to satisfy the levitation
condition for which the reference WP is stationary in the laboratory frame (i.e., relative to Earth).
In case of small experimental errors, one can consider the general case that includes deviations from the levitation condition, for which a simple calculation of the actions along the two interferometer trajectories when the closing condition is satisfied results in a measured QGI phase
\begin{equation} \Delta\phi=\frac{ma(a-2g)}{3\hbar}T^3\,,
\end{equation}
such that the outcome of the QGI depends on both the applied external acceleration on the reference arm and the free-fall acceleration, and becomes equal to the result of Eq.\,(2) in the case $a=g$, with a quadratic deviation $m(a-g)^2T^3/3\hbar$ if $a\neq g$. While the measurement of $g$ is not the aim of our experiment, we should note that the sensitivity with which a $T^3$ SGI, such as ours, can measure $g$ was recently estimated\,\cite{Zuniga2024}.

The momentum splitting of the WPs in the experiment requires a finite kick duration $T_{kick}$ in which the reference WP is temporarily in state $|0\rangle$ and the ballistic WP is temporarily in state $|1\rangle$ allowing it to be accelerated with an average acceleration $a_{kick}=v_0/T_{kick}$. Furthermore, in order to perform the internal-state flip before or after the kick, a finite delay time $T_d$ is needed between the kick pulse and the holding pulse of duration $T_h = 2T-2T_d$, where 2T represents the free-fall time of the ballistic WP. As explained in Methods, the ballistic WP has a free-fall time of $2T$, while the reference WP is in free fall during the time $T_{kick}+T_d$ before and after the holding time. Using $\phi_{gauge}$ these latter parts of the reference WP trajectory involve a free-fall phase in the laboratory (Newtonian) frame of
\begin{equation}
\phi_{\rm reference}=\frac{2m}{3\hbar} g^2(T_{kick}+T_d)^3\,. 
\end{equation}
Consequently, the phase of the reference WP is independent of the holding time.

By using arguments similar to the ones used above to derive the previous equations, we can also calculate the phase of the ballistic WP during the kick time and obtain the total phase difference expected in the experiment (SM)
\begin{equation}\label{eq:phase_diff}
\Delta\phi = \frac{mg^2}{3\hbar}\left(T^{3} + T^{2} T_{kick} + T \left(T_{kick}^{2} + T_{kick} T_{d}\right) - T_{d} \left(T_{kick} + T_{d}\right)^{2}\right)\,,
\end{equation}
where the quadratic and linear dependence on the interferometer time $T$ is due to the finite duration of the kick $T_{kick}$ (setting $T_{kick} = 0$ eliminates them), while $T_d$ adds a constant phase term.

Let us briefly describe in more detail the experimental sequence we have realized (more details in Methods and SM): We create a Bose-Einstein condensate (BEC) consisting of about $2\cdot10^4$ atoms in a magnetic trap, utilizing current-carrying wires on an atom chip (the atoms are below the chip which is upside down [Fig.\,1(B)]). We then release the atom cloud from the trap and conduct a Delta-kick cooling (collimation) procedure. The atom cloud is by now dilute enough so that atom-atom interaction is minimized and the physics becomes essentially single-particle physics. After a $\pi/2$ pulse, each atom is in a superposition of a magnetic sensitive state $|1\rangle$ and a non-sensitive state $|0\rangle$.

Next, we apply a magnetic kick which launches state $|1\rangle$ upwards. Quickly thereafter, we apply a $\pi$ pulse, so that the WP launched upwards is in state $|0\rangle$ and is no longer affected by the magnetic gradients, thus going into a ballistic trajectory affected only by gravity (up to second order Zeeman).

Conversely, the other WP, which is now in state $|1\rangle$, is sensitive to the magnetic gradient which is at this time tuned so that it exactly counteracts gravity. Consequently, this WP remains static and does not change position. Finally, we reverse the sequence to bring the two WPs to have the same position and momentum for a full overlap.

The detection stage starts with a $\pi/2$ pulse which projects the phase difference between the spin states to the population of the states. Next, another magnetic gradient kick separates the two spin states in
space, so that after some time of flight we are able to determine the population in each spin state using absorption imaging with a CCD, and thus measure the output of the interferometer.

Beyond the achievement of two very different WP trajectories in space, a significant experimental challenge lies in the fact that the two trajectories are very much different in momentum, and recombining the WPs in order to observe an interference pattern requires a good overlap in both position and momentum (namely, zeroing any difference, or in other words, time-reversing the splitting process). The time-irreversibility of the splitting process, whether it be of technical or fundamental origin, has been termed by Scully, Englert and Schwinger the ``Humpty-Dumpty effect" and combating it requires precise control over the trajectories\,\cite{margalit2021}.

More so, for a good overlap the size and shape of the two WPs have to be quite similar, and as they are exposed to different magnetic gradients having a different curvature, lensing effects make such an overlap hard to achieve. Finally, also rotation of the WPs in 3D may diminish the eventual overlap, and consequently the contrast (visibility) of the interference pattern\,\cite{japha2023}.

Whereas recombination in atom interferometers based on laser pulses, even when the momentum splitting is large\,\cite{Hammerer2023}, enjoys the naturally given quantum accuracy of the photon momentum, in an SGI the momentum transfer is classical in nature. This requires a high level of control over the magnetic fields as well as a detailed simulation of the WP evolution when exposed to such fields and gravity\,\cite{japha2021}.

\section{Results}

In Fig.\,2 we present the raw data, where we observe about 13 oscillations, which constitute about 80\,rad phase accumulation. In Fig.\,3 we analyze the phase extracted from Fig.\,2, and compare it to a numerical simulation, the analytical prediction of Eq.\,(\ref{eq:phase_diff}) as well as to the {\it Gedanken} experiment [Eq.\,(2)]. The analytical approximation assumes square pulses of a homogeneous gradient along the $z$-axis giving rise to a purely one-dimensional motion along $z$, which together with the symmetry of the scheme results in perfect path recombination. The numerical simulation takes into account (i) three-dimensional propagation in a curved potential resulting in a non-perfect path recombination, (ii) internal WP dynamics (expansion, focusing and rotation) and (iii) atom-atom interactions within each WP. The first two factors may change the phase by an order of 1-2 radians each, while the latter may contribute up to about 0.25 radians. Obviously, the prediction of the numerical simulation is accurate to within the uncertainty limits of the experimental parameters. As can be seen from Fig.\,3(B), the residuals amount to about 2.5\%, presenting good agreement between the model and the data for the $t^3$ term. Finally, and importantly (see introduction), the data is also consistent with the predicted value of the $mg^2/6\hbar$ prefactor, as shown in Fig.\,3(C).

In the SM, we provide detailed information as to how all lines in Figs.\,2 and 3 were plotted.
The main source of uncertainty in the experimental results arises from the systematic uncertainty in the ratio of the kick and holding currents, as the phase accumulation is most sensitive to this ratio, where deviations from the desired ratio add a phase term that scales linearly with the interferometer time.

\begin{figure}[H]\label{fig:population_vs_T.pdf}
\centering
\includegraphics[width=1\textwidth]{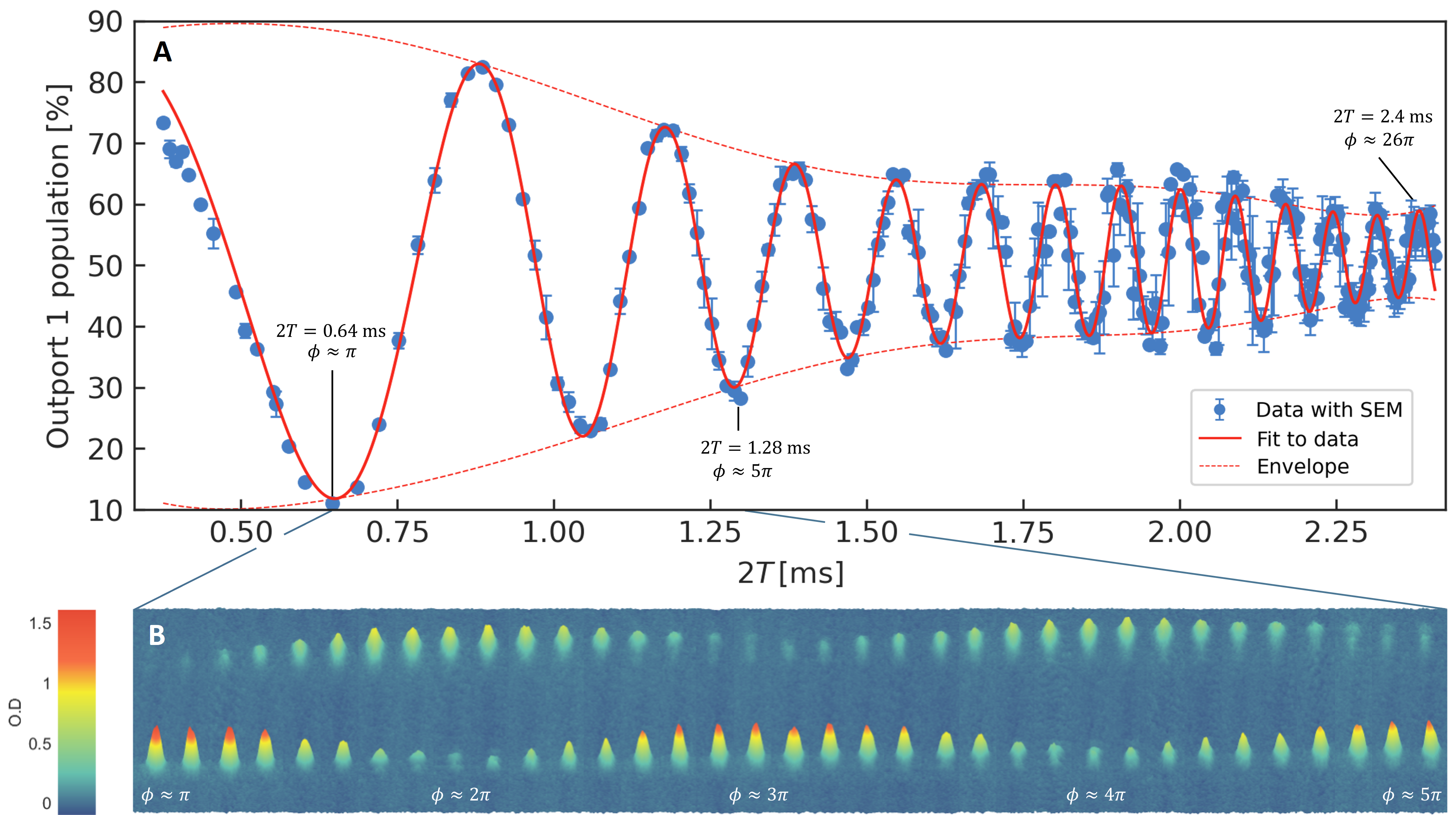}
\caption{\small
{\textbf{Population in outport 1 vs. the free-fall duration of the ballistic path ${2T}$.}} \newline (\textbf{A}) The relative spin population in outport 1 (state $|0\rangle$) oscillates with a distinct chirp as a function of interferometer time. The blue points represent the experimental data, with error bars derived from the standard error of the mean (SEM). The red line is a fit to the data of the form $P = P_{\text{mean}}(2T) + \frac{1}{2}V(2T)\cos\left[\phi(2T)\right]$, where $V$ is the visibility of the oscillations, $\phi$ is the phase of the oscillations and $P_{\text{mean}}$ is the mean value of the population. To obtain the fit, we first identify the upper and lower envelopes and fit them to a polynomial to get $P_{\text{mean}}$ and $V(2T)$. We then fit the data to the model where the phase $\phi(2T)$ is a third-order polynomial. During the fitting we exclude the first and last oscillation as they suffer from larger uncertainty in the values of the envelope. The data shows 13 complete oscillations, with low phase noise, resulting in an average population error (SEM) of $1.8 \pm 1.5$ [\%] and good visibility, starting at $V=80\,\%$ and decreasing to $V=20\,\%$ at $2T = 2000 \, \mu\mathrm{s}$. The decrease in visibility is mainly due to a varying overlap between the WPs due to differences in shape evolution under the influence of curved gradients as well as imperfection in the path recombination. The figure includes $633$ experimental cycles, each $30$\,s, so the graph was taken over a period of $5.3$ hours. The short-term and long-term stability of the interferometer are good due to the high stability of the chip trap, and current source. (\textbf{B}) Atomic distribution (heat map of optical density) adapted from CCD images (on-resonance absorption imaging), of the atoms in the two interferometer outports [see Fig.\,1(B)], showing two oscillations. The top cloud is outport 1 (state $|0\rangle$), and the bottom cloud is outport 2 (state $|1\rangle$, which is magnetically sensitive). The cloud in outport 2 is focused by the final splitting pulse, increasing its optical density, and making it appear slightly denser even when the two clouds have the same population. The vertical scale of the image is 1.1\,mm.}
\end{figure}

\begin{figure}[H]\label{fig:Phase}
\centering
\includegraphics[width=0.9\textwidth]{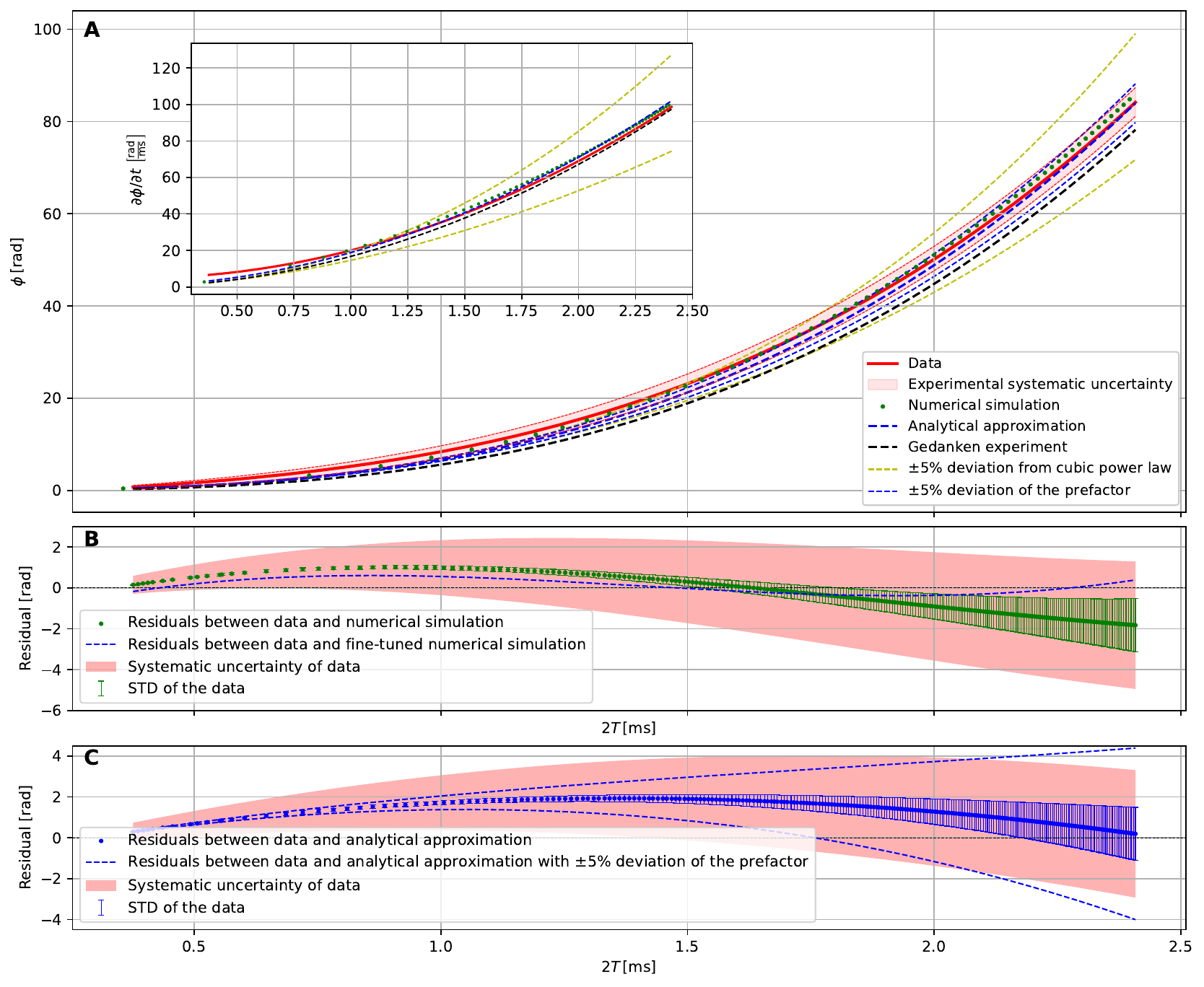}
\caption{\scriptsize
{\textbf{Phase and its derivative vs. the free-fall duration of the ballistic path {$2T$}.
}}
\newline(\textbf{A}) The phase difference between the two interferometer arms as a function of the free-fall duration of the upper arm. The red line represents the experimental data, where the phase is extracted from Fig.\,2. To estimate the uncertainty bounds, we repeat the experiment with an increase (decrease) of the magnetic-pulse current $I_{kick}$ by $0.5\,\%$ (which is our estimated experimental precision) and use the result to plot the upper (lower) uncertainty bounds. The green points are the result of a numerical simulation based on tools we developed for calculating WP evolution\,\cite{japha2021}, which include the effects of shape and rotation of the WP. The blue dashed line is the result of the analytical approximation of the experiment given in Eq.\,\ref{eq:phase_diff}. The analytical model assumes a homogeneous magnetic gradient, and a set of square pulses of current so that the acceleration of each arm is a piecewise constant function. The black dashed line depicts the phase of the \textit{Gedanken} experiment. The inset shows the derivative of the phase with respect to $2T$. To obtain the derivative of the phase for the numeric simulation, we fit a third-order polynomial to the set of points in the main figure and take its derivative. The derivative of the analytical approximation and the \textit{Gedanken} experiment are calculated by direct differentiation. The yellow dashed lines represent 5\% deviations from the $T^3$ phase accumulation model, by modifying the analytical prediction to a power law of $T^{3\pm0.15}$. The thin blue dashed lines represent the analytical approximation with 5\% deviations from the theoretical prefactor (note that the prefactor in our experiment is $mg^2/3\hbar$, as explained in Eq.\,\ref{eq:ballistic}). This is meant to show that the data excludes such deviations from the cubic phase or the predicted prefactor. (\textbf{B}) The residuals between the phase of the data and the numerical simulation, where the latter is a ``blind" simulation with no free parameters or fine tuning. A difference of 2\,rad over 13 oscillations spanning a phase of $\approx 80$\,rad, constitutes a deviation of about 2.5\%. The error bars represent the statistical noise of the phase in the experiment, which shows a relative noise of 1.3\% of the phase. The red band represents the systematic uncertainties [which also appear in (\textbf{A})]. The blue line shows the result of the numeric simulation after fine-tuning of the currents and the magnetic field along the z-axis. The kick current is increased by 0.15\% , the idle current is changed from 0.55\,mA to 0.5\,mA, and $B_z$ from 0.33\,G to 0.35\,G, all within the experimental uncertainties. (\textbf{C}) The residuals between the phase of the data and the analytical approximation. The 5\% lines allow one to put an upper limit of a few percent on the possible deviation from the predicted prefactor.}
\end{figure}

\section{Discussion}

The good agreement between the data and the analytical as well as numerical prediction, brings us to conclude that we have indeed observed the expected interferometer phase difference of $-\frac{m}{3\hbar}g^2t^3$ [Eq.\,(2)] and in doing so, we have confirmed with a high level of confidence the predicted $-\frac{m}{6\hbar}g^2t^3$-phase [Eq.\,(1)]. We have thus observed the phase of a freely-falling object.

Although measuring $g$ is clearly not the goal of our experiment, the phase of a freely-falling object is proportional to $g$ so the measurement of the phase may be considered a measurement of gravity. Measuring gravity has been done for hundreds of years, and ever since the celebrated COW experiments \,\cite{Colella1975,Rauch2015}, also with matter-wave interferometers. Finally, for the last three decades, measuring gravity has been done also with atom interferometers such as ours\,\cite{Chu0,Tino2021,Lellouch2022,Chu1,Lin2011,ChenYang2025,Cassens2025}, and it has already become part of technology\,\cite{Wu2019,Antoni‐Micollier2022,Janvier2022,Fenbg2024}, but one should nevertheless acknowledge subtleties. Specifically, it is agreed that the EP in fact indicates that all tests of fundamental physics (including gravitational physics) are not affected, locally, by the presence of a gravitational field\,\cite{Casola2015}. One may interpret this to mean that matter-wave interferometers cannot measure gravity in the absolute sense in a uniform gravitational field but it is understood that if a part of the apparatus is firmly connected to Earth (the source mass) then this constitutes a non-local measurement and gravity may be measured\,\cite{Asenbaum2024}.

In the QGI this non-local information is transferred from the apparatus to the reference WP, by the levitation condition. More explicitly, the position of the reference WP, as well as the apparatus, is fixed relative to the surface of Earth, located at a distance $r=R$ from the center of Earth ($r=0$), and hence, ignoring rotations and tidal movements, has the same acceleration as the center of Earth, $g(r=0)$, which may have a nonzero value due to some external gravitational field (e.g., from the Sun). The free-falling WP experiences $g(R)$, which in addition to $g(0)$ is due to the gravitational pull of Earth, again ignoring rotations and tidal movements. The relative acceleration of the two WPs is thus determined by the gravity of Earth.
 
The QGI has a unique geometry which allows to directly measure the phase of a free-falling WP due to gravity. Obviously, observing from the Einsteinian frame, the observer may claim that the phase is due to the magnetic acceleration, but this is of course also a measure of gravity due to the levitation
condition, which determines the equality:
\begin{equation}
\Delta\phi= -\frac{(mg)^2T^3}{3\hbar m}=-\frac{F_{\text{mag}}^2T^3}{3\hbar m}\,.
\end{equation}

To conclude, we have measured for the first time the quantum phase of a freely-falling object, and at the same time we have confirmed the phase predicted if the EP is applied to a freely-falling WP. Nature, it seems, at these low masses and energies, does indeed accommodate both QM and the EP in this delicate manner. In the outlook, we look beyond the present work and consider how future versions of the experiment may draw us into even deeper waters. In this light, the present experiment may be seen not merely as a confirmation of a long-predicted quantum phase, but as a potential stepping stone toward a deeper synthesis of gravity and quantum theory.

\section{Outlook}

As an outlook, let us introduce the term -- quantum equivalence principle (QEP). As there seems to be no agreement on how to exactly define the QEP, we leave this for future theoretical and experimental work\,\cite{marletto2020,Lammerzahl1996,Rosu1999,Padmanabhan2011,Pereira2015,Zych2018,Brukner2019,HARDY2019,Das2023,Brukner2022}. Indeed, quite a few works claim that the $t^3$ phase is also a fundamental aspect of the EP for quantum systems\,\cite{Nauenberg2016,Greenberger1979,marletto2020,Brukner2019}. Hence it may be argued that the QGI experiment described here also observed a consequence of the QEP. Other, more complex tests, are also being suggested\,\cite{Geiger2018}.

Let us note two final points: First, as the QGI is also able to manipulate clock WPs (for clocks in an SGI, see\,\cite{Margalit2015}), it opens the door for the test of the more complex formulations suggested for the EP in the quantum domain. Such an interferometer also opens the door to searches for new physics with clocks\,\cite{Schleich2013,Loriani2019,Giese2019,Roura2020,pumpo2023}. Second, such an interferometer is a milestone towards a SGI with nanodiamonds instead of atoms, for a test of quantum mechanics in new regimes and to probe the quantum-gravity interface\,\cite{Marletto2025}, e.g., test the quantization of gravity\,\cite{margalit2021, Bose2017, Marletto2017}. In addition, such a large-mass (active-mass) interferometer can explore the D\'iosi-Penrose conjecture regarding gravitationally-induced collapse, and possible tensions which may arise with the QEP\,\cite{RevModPhys.29.423,Bassi2013,RevModPhys.97.015003}\,(see\,SM).

\section{Acknowledgments}
We thank the BGU nano-fabrication facility for the high-quality chip and the BGU support team, especially Menachem Givon, Zina Binstock, Dmitrii Kapusta and Yaniv Bar-Haim for their support in building and maintaining the experiment. Funding: This work was funded, in part, by the
Israel Science Foundation (grants no. 856/18, 1314/19, 3515/20, and 3470/21), and the German-Israeli DIP project (Hybrid devices: FO 703/2-1) supported by the DFG. This work has been supported by the "Table-top experiments for fundamental physics" program, sponsored by the Gordon and Betty Moore Foundation, Simons Foundation, Alfred P. Sloan Foundation, and John Templeton Foundation. I.F. thanks an anonymous USA philanthropist, John Moussouris, Jussi  Westergren and the Emmy Network for support and research funding. C.M. and V.V. thank the Gordon and Betty Moore Foundation and the Templeton Foundation and C.M. also thanks the Eutopia Foundation, and Ben Vass for supporting her research. W.P.S. is grateful to the Hagler Institute for Advanced Study at Texas A$\&$M University for a Faculty Fellowship, and to Texas A$\&$M AgriLife Research for the support of this work. We thank Časlav Brukner, Daniel Rohrlich, Daniel M. Greenberger, Ernst M. Rasel, Yair Margalit, Jannik Str\"ohle, Gary G. Rozenman, Fabio Di Pumpo and Matthias Zimmermann for helpful discussions.

\section*{Methods}

\begin{figure}[H]
\centering
\includegraphics[width=1\textwidth]{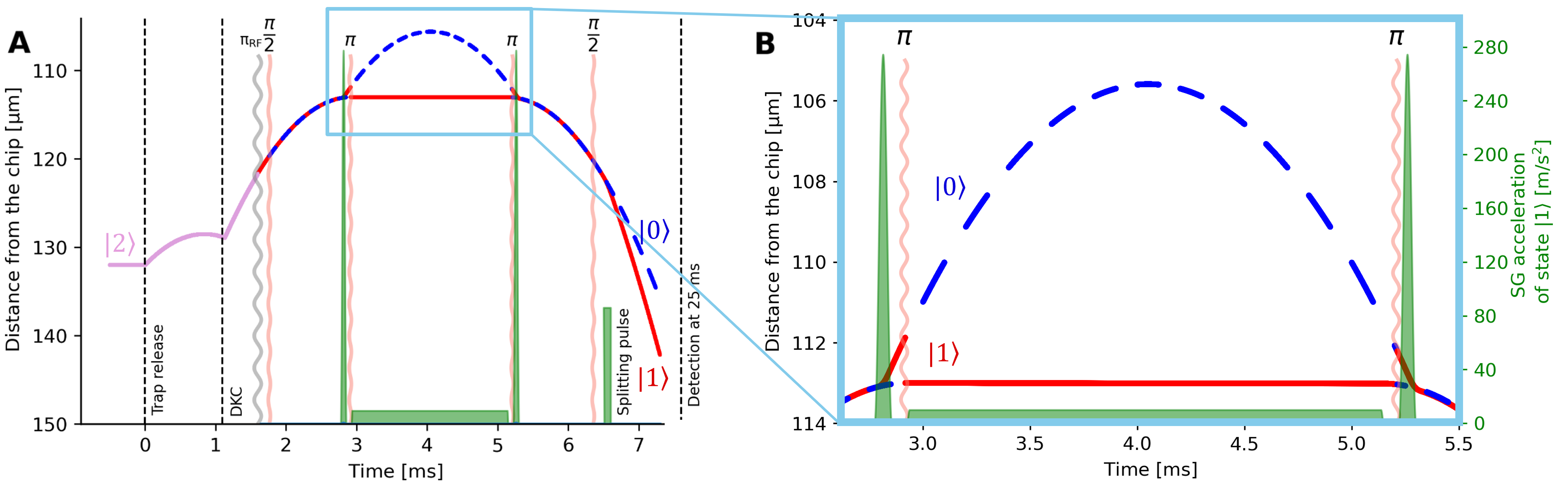}
\caption{\footnotesize
{\textbf{The experimental scheme and trajectories of the WPs in the experiment.
}}
\newline
(\textbf{A}) The complete experimental scheme of our longitudinal (1D) QGI, for the case of $2T = 2.4\,\rm ms$. The x-axis is time, where we set $t=0$ at the moment of trap release. At this time the atoms are in the state $|2\rangle \equiv|F=2,m_F=2\rangle$. We apply Delta-kick cooling (DKC) at $t=1.1\,\rm ms$, which collimates the WP's expansion and launches the atoms into a ballistic trajectory upwards. Before the start of the interferometer, we transfer the atoms to the state $|F=2,m_F=1\rangle$ by applying an on-resonance RF $\pi$ pulse (grey wavy line). During the interferometer, we control the spin state of the atoms with four MW pulses (orange wavy lines) and the momentum of the atoms with three magnetic gradient pulses (green areas). The MW pulses are on resonance with the transition between the $|1\rangle \equiv|F=2,m_F=1\rangle$ and $|0\rangle \equiv|F=1,m_F=0\rangle$ states, where the $|0\rangle$ state is magnetically non-sensitive in the Zeeman first order. The first $\pi/2$ pulse puts the atoms in an equal superposition of $|1\rangle + |0\rangle$. At the end of the interferometer, the trajectories of these two states are joined to a single trajectory with two spin states, after which the second $\pi/2$ pulse is applied, creating four overlapping WPs which interfere, two WPs per each spin state, imprinting the phase difference into a population difference (for simplicity of graphics, we have not differentiated the four WPs in the plot). After the second $\pi/2$ pulse, we detect the relative population in each state. In the detection scheme, we first spatially separate the states by applying a fourth magnetic gradient and, finally, image the atoms using on-resonance absorption imaging. The symmetric temporal positioning of the $\pi/2$ and $\pi$ pulses also serves as a dynamic-decoupling scheme, increasing the coherence time, canceling the linear phase accumulation due to possible detuning of the MW and reducing the shot-to-shot phase fluctuations.
(\textbf{B}) Zoom-in on the interferometer sequence. The first gradient (kick) pulse applies a maximal acceleration $a_{kick}$ for a total duration of $T_{kick}$, which launches the top arm into a ballistic trajectory. $50\,\rm \mu s $ after the end of the gradient pulse, we apply a $\pi$ pulse (with a duration of $16\,\rm \mu s$) that inverts the spin state of the arms so that the bottom arm is now in the magnetically-sensitive state $|1\rangle$. $5\rm\,\mu s$ after the $\pi$ pulse, we start the holding pulse (duration of $T_h$ and rise time $\tau_h = 12\,\rm \mu s $), which applies an acceleration opposite and equal to gravity, to hold the bottom arm stationary in the laboratory frame. $50\,\rm \mu s $ after the holding pulse, we again apply a $\pi$ pulse and $5\rm\,\mu s$ later we recombine the trajectories of the two WPs by applying a magnetic gradient pulse with a maximal acceleration $a_{kick}$ for a total duration of $T_{kick}$, as in the first pulse.
}
\label{fig:Scheme_V1.png}
\end{figure}

Let us describe the experimental sequence in some detail (see Fig.~\ref{fig:Scheme_V1.png}): we create a $^{87}$Rb BEC in the state $|2\rangle \equiv|F=2,m_F=2\rangle$ in a magnetic trap produced by current-carrying wires on the atom chip (below the chip which is upside down). We then release the cloud of atoms from the trap and conduct Delta-kick cooling (DKC), which collimates the expansion of the WP along the z axis down to an effective temperature of 3.3\,nK. As a side effect of this procedure, the atoms are launched upwards (as can be seen on the left side of Fig.~\ref{fig:Scheme_V1.png}). The BEC is by now dilute enough so that atom-atom interaction is minimized and the physics becomes single-atom physics. We then apply $100\,\mu$s radio-frequency (RF) $\pi$ pulse transferring atoms to the state $|F=2,m_F=1\rangle$. With the help of a microwave (MW)  $\pi/2$ pulse, each atom is now in a superposition of a magnetic sensitive state $|1\rangle \equiv|F=2,m_F=1\rangle$ and a non-sensitive state $|0\rangle \equiv|F=1,m_F=0\rangle$. We now apply a magnetic SG kick which launches the sensitive state upwards. Quickly thereafter, we apply a MW $\pi$ pulse, so that the atoms launched upwards are no longer affected by the magnetic gradients and they go into a ballistic trajectory affected only by gravity (up to second order Zeeman). Conversely, the other WP becomes sensitive to the magnetic gradient which is now tuned to exactly counteract gravity, so that the WP remains stationary in the frame of reference of the laboratory. Finally, we reverse the sequence -- apply a MW $\pi$ pulse and apply the same SG kick to stop the ballistic WP, thus bringing the two WPs to have the same position and momentum for a full overlap. After one more MW $\pi/2$ pulse, another magnetic gradient kick separates the two spin states in space, so that after some time of flight we are able to determine the population in each output port and consequently in each spin state, and thus measure the output of the interferometer (Fig.\,2).

The SG force is created by currents in three parallel gold micro-wires of $2\,\mu$m thickness, $40\,\mu$m width, separated by $60\,\mu$m gaps on the atom chip surface. The current direction in the middle wire is opposite to that in the edge wires, which creates a quadrupole magnetic field with a minimum at approximately $100\,\mu$m from the wires. The external magnetic bias field of 12.6\,G is applied to suppress spin-flips and introduce a small second-order Zeeman shift for the efficiency of the RF pulse.

The optimal value of the holding current $I_{\rm hold}$ should counteract gravity. To find $I_{\rm hold}$, we measured trajectories of state $|1\rangle$  with resonant absorption imaging for various values of the holding current (see SM for details) and find $I_{\rm hold} = 23.07 \pm 1.49$\,mA. Although the inaccuracy of gravity compensation is 6.5\%, it corresponds to only 0.42\% inaccuracy in the phase accumulation, as the deviation from $g$ is introduced squared to the phase (see SM). We note that the phase of Eq.\,(3) has a maximum value at $a=g$ which reduces the sensitivity to small changes of the magnetic acceleration, while keeping the sensitivity to the gravitational acceleration linear.

\begin{figure}[H]
\centering
\includegraphics[width=1\textwidth]{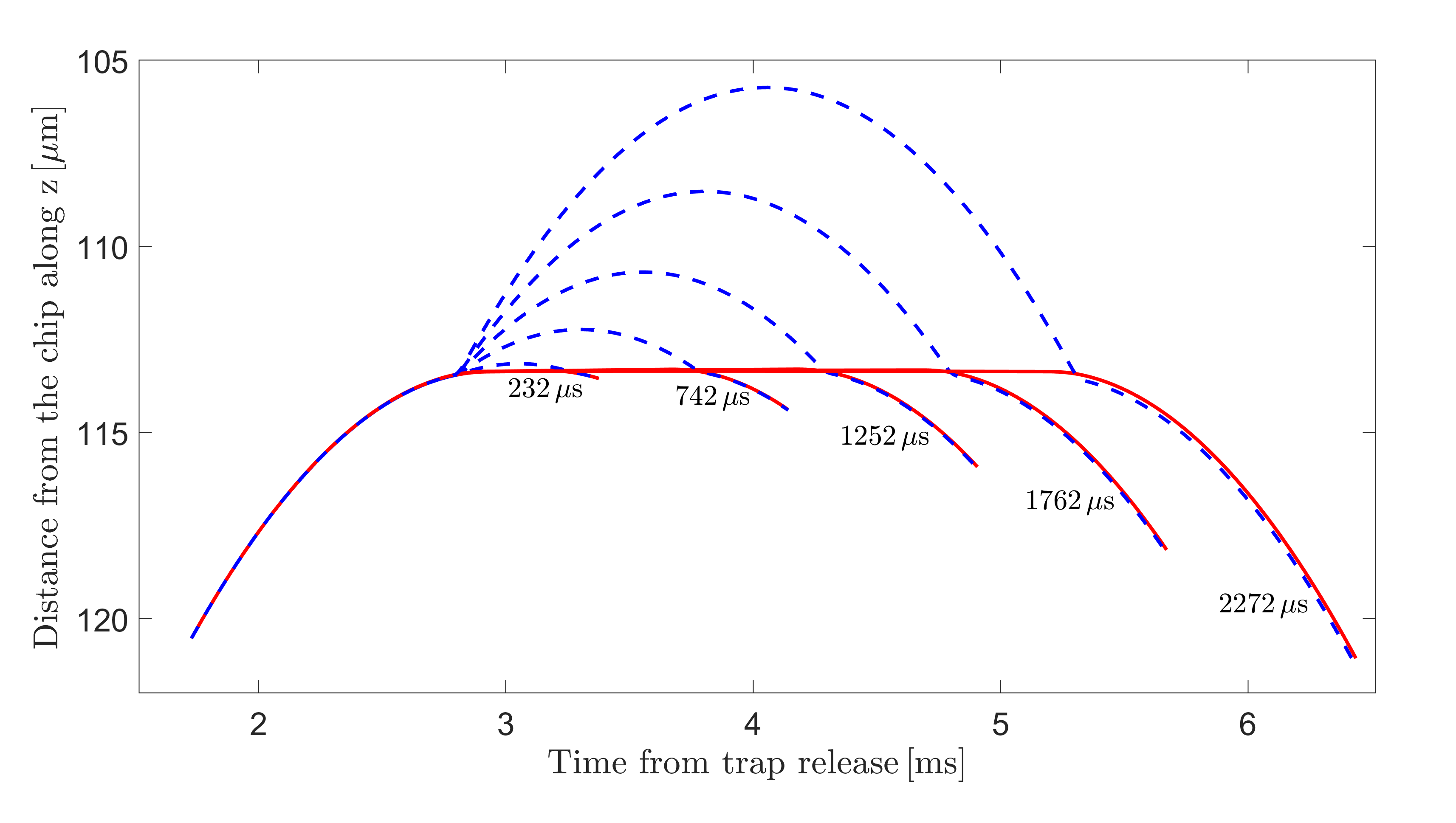}
\caption{
{\textbf{The trajectories of the WPs for different holding durations.
}}
\newline
The trajectories of the two arms of the interferometer, calculated by the numerical simulation for holding durations $T_h = 232, 742, 1252, 1762, 2272 \,\rm \mu$s. Dashed blue lines represent the ballistic trajectory, and solid red lines the static WP. At the end of the interferometer, a small spatial splitting can be seen, on the order of $0.1 \,\rm \mu m$. For $T_h = 2272 \,\rm \mu$s, the spatial splitting between the two arms halfway through the interferometer reaches $7.5 \,\rm \mu$m. The trajectories are plotted until the time in which the final (second) $\pi/2$ is applied.
}
\label{fig:Trajectories_V1.png}
\end{figure}


The kick currents are set in such a way that the area $\int I(t) dt$ of the two kick current pulses is equal to the area of the holding current pulse for each interferometer duration. In the case of the {\it Gedanken} experiment and the analytical approximation, such a definition of the kick pulses is equivalent to the condition of full overlap between the WPs at the end of the interferometer. However, in the real experiment, this condition does not result in perfect overlap due to the curvature of the potential and the finite duration of the pulses. This is quantified in the numerical simulation, where the precise pulse area condition is applied, the two trajectories do not completely overlap at the end, resulting in a small final spatial splitting of $0.2\,\rm{\mu m}$ for the largest interferometer duration (as seen in Fig.~\ref{fig:Trajectories_V1.png}). The effects of rotations and the expansion of the WPs under the curved potential are also quantified in the numerical simulation and contribute up to 1\,rad to the total phase (as seen in Fig.\,\ref{fig:Numerical_simulation_phase_trems.pdf}). While in Fig.~\ref{fig:Scheme_V1.png} the trajectories produced by an analytical approximation, in Fig.~\ref{fig:Trajectories_V1.png} we present several trajectories as they are calculated by the numerical simulation, and as in Fig.\,3, the two are in good agreement.

The maximal reached spatial splitting between the two arms is approximately $7.5\,\mu$m, based on the numerical simulation and in agreement with the analytical estimations and trajectory imaging.

Next, let us briefly describe the timings of the MW pulses. We use a symmetric dynamical decoupling MW scheme, as it cancels the linear phase due to possible detuning of the MW pulses, improves the coherence time, and reduces the shot-to-shot noise. The rise and fall time of the holding pulse (duration of $T_h$) are each defined as $\tau_h = 12\,\rm \mu s $. The time between the first and second $\pi$ pulses defines the MW dynamical decoupling scheme. We define it as $2\tau_{\rm DD}$, which amounts to $2\tau_{\rm DD} = 5+2\times \tau_h+T_h+50$, where $5\rm\,\mu s$ after the $\pi$ pulse we start the holding pulse, and $50\,\rm \mu s $ after the end of the holding pulse we apply the second $\pi$ pulse (Fig.~\ref{fig:Scheme_V1.png}). We then set the time between the $\pi/2$ and $\pi$ pulses to be $\tau_{\rm DD}$.

Finally, let us describe with a simplified calculation (details in the SM) how the $T_h = 2272 \,\rm \mu$s (Fig.~\ref{fig:Trajectories_V1.png}) or the $2T = 2.4\,\rm ms$ (Fig.~\ref{fig:Scheme_V1.png}) achieve a splitting of about $7.5\,\rm \mu$m. We define $T_d$ as the delay between the end of the gradient (kick) pulse and the start of the holding pulse, where $50\,\rm \mu s $ after the end of the gradient pulse we apply a $\pi$ pulse (with a duration of $16\,\rm \mu s$) that inverts the spin state of the arms so that the bottom arm is now in the magnetically-sensitive state $|1\rangle$, and $5\rm\,\mu s$ after the $\pi$ pulse we start the holding pulse. This amounts to $T_d = 50+16+5 = 71\,\rm \mu s$, where the same delay time appears also after the holding pulse.
The top arm travels $0.5\times g_{eff}\times (T_{eff})^2 = 7.62\,\mu$m, where $g_{eff} =9.91$\,m/$s^2$ (including acceleration due to the second order Zeeman shift), and $2T_{eff} = T_h+2\times \tau_h+2\times T_d +T_{kick} =2T+80\,\mu$s, is the effective ballistic duration taking into account the translation along z during $T_{kick}$ (only one $T_{kick}=80\,\mu$s is added as an average of the two kicks).
The bottom arm travels upwards $0.5\times g_{eff}\times (T_{kick}+T_d)^2 = 0.11\,\mu$m during $T_{kick}$ and $T_d$, resulting in a maximal splitting of $7.5\,\mu$m. This is in good agreement with the numerical results presented in Fig.~\ref{fig:Trajectories_V1.png}.

The complete details concerning the experiment may be found in the SM.


\clearpage 

%

%
%
%
%
%
%




\newpage


\renewcommand{\thefigure}{S\arabic{figure}}
\renewcommand{\thetable}{S\arabic{table}}
\renewcommand{\theequation}{S\arabic{equation}}
\renewcommand{\thepage}{S\arabic{page}}
\setcounter{figure}{0}
\setcounter{table}{0}
\setcounter{equation}{0}
\setcounter{page}{1} 


\begin{center}
\section*{Supplementary Materials for\\ \scititle}

  Or Dobkowski, Barak Trok, Peter Skakunenko, \and
  Yonathan Japha, David Groswasser, Maxim Efremov,
  Chiara Marletto, \and Ivette Fuentes Guridi, 
  Roger Penrose, Vlatko Vedral, \and Wolfgang P. Schleich, Ron Folman
\end{center}


\newpage


\renewcommand{\thesubsection}{\Alph{subsection}}

\subsection{BEC production}
\begin{figure}[H]
    \centering
    \includegraphics[width=0.9\textwidth]{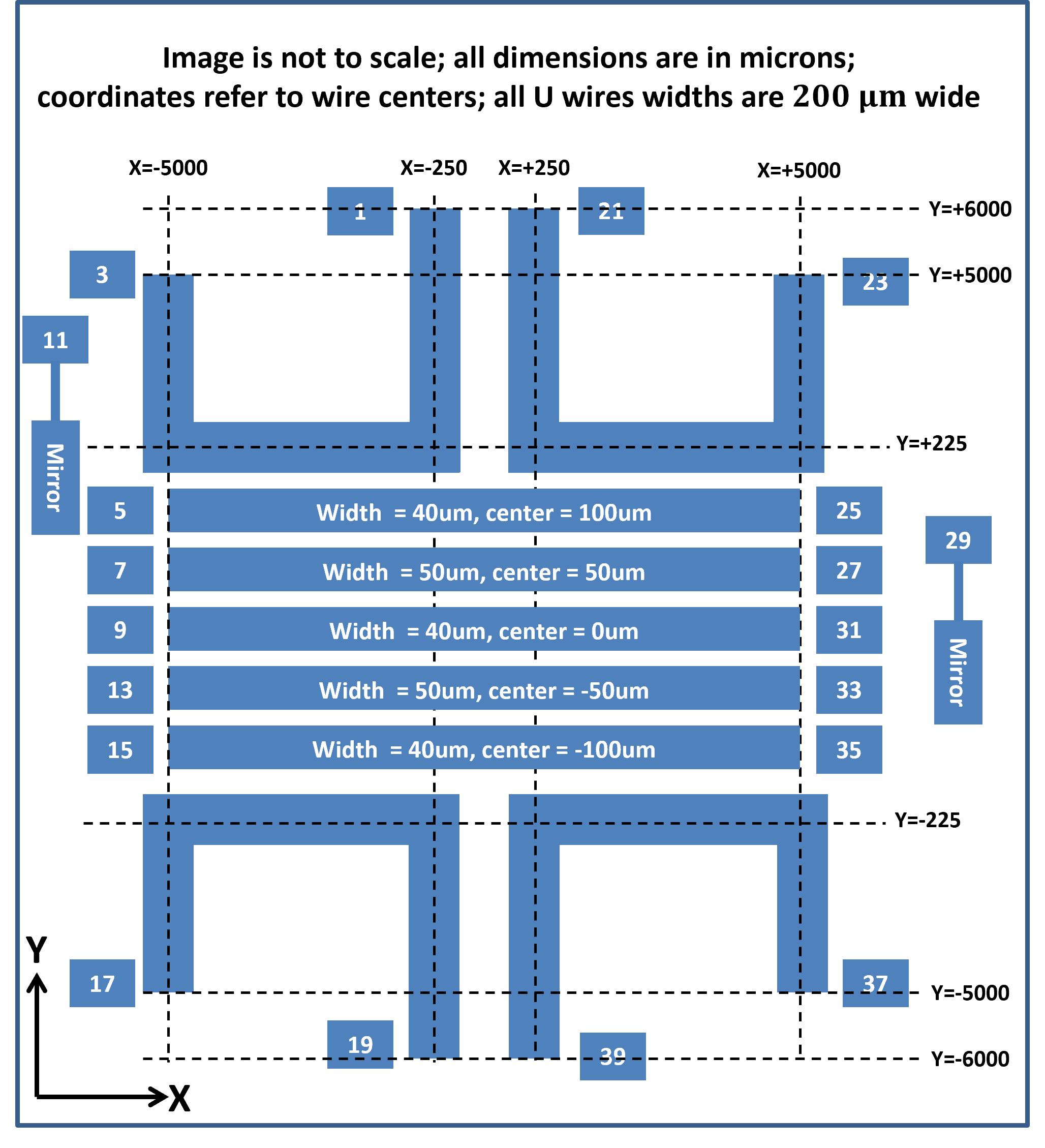}
    \caption[Atom chip wires layout]{Schematic representation of the wires on the atom chip. The thickness of the wires is $2\rm\,\mu m$, and their width ranges from $40-200\rm\,\mu m$. At the center of the chip are five straight wires. The three SG wires have a width of $40\rm\,\mu m$ each, and their centers are separated by $100\rm\,\mu m$ (labeled 5, 9, 15) with alternating current polarity. The SG wires produce a quadrupole $97\rm\,\mu m$ below the chip surface. The magnetic trap on the chip consists of two straight wires and four U-shaped wires. The straight wires (labeled 7, 13) confine the atoms in the z and y axes and have a width of $50\rm\,\mu m$, and their centers are separated by $100\rm \,\mu m$. The four U-shaped wires confine the atoms along the x-axis and have a width of $200\rm\,\mu m$ each. This confinement is produced by the inner wires (labeled 1, 19, 21, 39).}
    \label{fig:BGU2 chip wires layout.png}
\end{figure}

We produce a Bose-Einstein condensate (BEC) of $\rm Rb^{87}$ atoms with $15-30\cdot10^3$ atoms in the condensate. To produce the BEC we load atoms from a 2D Magneto-Optical-Trap (MOT) to the 3D-MOT for 5 seconds, with an initial loading rate of $200\cdot 10^6$ atoms per second and a total of $600\cdot 10^6$ atoms in the MOT.
We compress the MOT and apply optical molasses, and optical pumping, after which we load the atoms to a magnetic trap formed by a copper z-wire. We apply evaporative cooling for 12 seconds and transfer the atoms to the chip trap, where we apply the final evaporation to a BEC for 1 second.
At this stage, the current in the y-bias coils is $19\rm\,A$, and in the x-bias coils $1\rm\,A$, and the magnetic fields are $12.6$\,G and $0.7$\,G, respectively. 
The chip trap consists of two straight wires with a current of $500\rm\,mA$ and four U-wires with a current of $700\rm\,mA$, the layout of the wires in the chip is shown in Fig.\,\ref{fig:BGU2 chip wires layout.png}. The trap frequency along the z axis was measured to be $1009\pm4 \rm\,Hz$, the magnetic field at the trap minimum induced a Zeeman shift of $425\pm5$\,kHz.

\subsection{Initial position stability}
\begin{figure}[h!]
\centering
\includegraphics[width=1\textwidth]{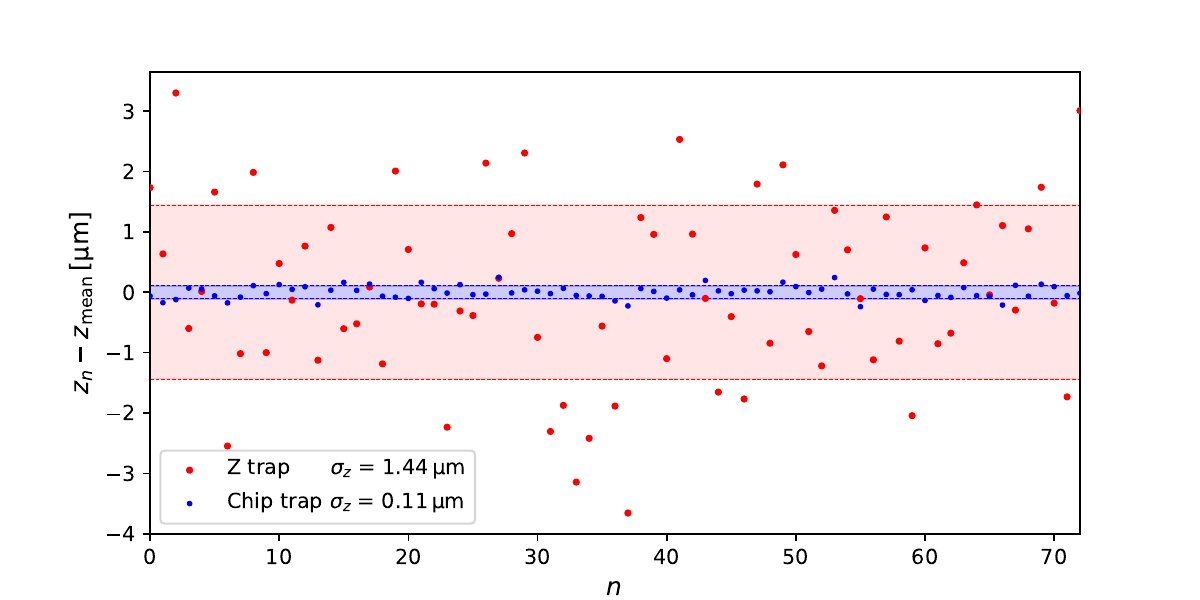}
\caption[Magnetic trap stability]{Initial position stability: Plotting the shot-to-shot fluctuations of the position of the atoms along the z-axis reveals a standard deviation of $0.11\,\rm\mu m$ in the atom chip trap, compared to $1.44\,\rm\mu m$ in the z-wire trap. The high stability of the atom chip trap significantly enhances the visibility and phase stability of the SGI.}
\label{fig:Trap stability}
\end{figure}

The phase and visibility of the Stern-Gerlach interferometer (SGI) are sensitive to the initial position of the atoms due to the curvature of the magnetic potential produced by the atom chip wires, thus the stability of the initial position is important both for the shot to shot stability and low frequency drifts. By loading the atoms to the atom chip trap we significantly improve the stability of the initial position compared with the macroscopic z-wire trap. To estimate the stability of the trap position, we measured the shot-to-shot fluctuations of the final position of the trap, resulting in a standard deviation of $0.11\rm\,\mu m$ (see Fig.\,\ref{fig:Trap stability}). The high positional stability is a key advantage of the atom chip trap over the z-wire trap, enabling improved visibility, enhanced phase stability, and greater spatial splitting compared to previous experiments \cite{Amit2019,margalit2021}. Using the atom chip trap also reduces low frequency noise (drifts) in the interferometer, enhancing the long-term stability of the phase, allowing for long sessions of data taking (3-6 hours of data for one graph), and high repeatability of the experiment.

\subsection{Trap release and Delta-kick cooling}
\begin{figure}[H]
    \centering
    \includegraphics[width=1\textwidth]{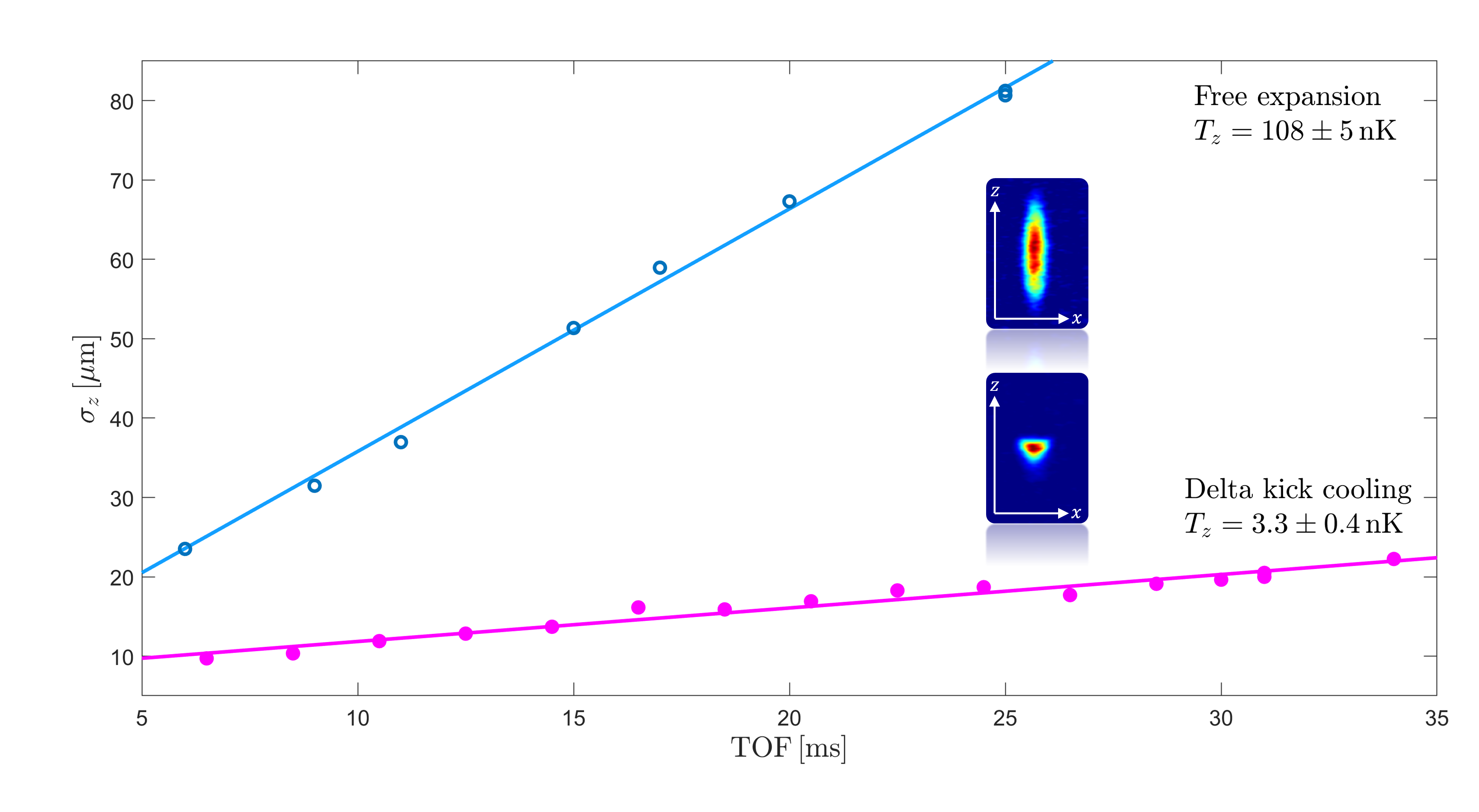}
    \caption[Temperature measurement along the $z$ axis, with and without Delta-kick cooling]{Temperature measurement along the $z$ axis, with and without DKC. The expansion rate with DKC is $0.4\,\rm\mu m/ms$, corresponding to a temperature of 3.3\,nK. For free expansion without DKC the expansion rate is $3.1\,\rm\mu m/ms$ corresponding to 108\,nK. The inset shows the CCD image of the atoms with and without DKC after a time of flight of 26.5\,ms and 25\,ms respectively.}
    \label{fig:DKC}
\end{figure}

We release the atoms from the trap by shutting off the current in the chip trap with a linear ramp of $30\rm\,\mu s$. The finite duration of the ramp gives the BEC some momentum upwards. After a time of flight of $1030\rm\,\mu s$, we apply Delta-kick cooling (DKC) by applying a pulse of current in the straight trap wires for $110\rm\,\mu s$ with a current of $470 \rm\,mA$, (the shape of the pulse is a trapezoid with linear rise of $9 \,\rm\mu s$ and linear fall of $22 \,\rm\mu s$).
The DKC collimates the BEC along the z and y axes and reduces its expansion rate to an effective temperature of $3.3\rm\, nK$ along the z axis (see Fig.\,\ref{fig:DKC} for a measurement of the temperature with and without DKC). A desired side effect of the DKC is a momentum kick to the BEC that launches it upwards toward the chip. Measurement of this trajectory reveals that the BEC reaches its maximum height $2930\pm 100\,\rm\mu s$ after the trap release. The peak of the trajectory is set to be the starting point for the holding pulse.

\subsection{Spin states, MW and RF pulses}
We use radio-frequency (RF) and micro-wave (MW) pulses to manipulate the spin state of the atoms. The atoms start in the state $|F=2,m_F=2\rangle$, and we apply an RF $\pi$ pulse, to move the atoms to $|1\rangle = |F=2,m_F=1\rangle$, with a frequency of $8.8\rm\,MHz$ and a duration of $100\rm\,\mu s$.
Our two-level system is
\begin{equation}
    |0\rangle=|F=1,m_F=0\rangle, |1\rangle=|F=2,m_F=1\rangle ,
\end{equation}
which we coherently control via MW pulses with a frequency of $6.843\rm\,GHz$, where the duration of a $\pi$ pulse is $16\rm\,\mu s$.
The MW scheme consists of four pulses arranged for dynamical decoupling (DD) to reduce environmental effects. The DD scheme consists of four MW pulses ($\pi/2,\pi,\pi,\pi/2$), with a delay of $\tau_{DD}$ between the $\pi/2$ and $\pi$ pulses, and $2\tau_{DD}$ between the two $\pi$ pulses, so that the scheme is ($\pi/2,\tau_{DD},\pi,2\tau_{DD},\pi,\tau_{DD},\pi/2$). This scheme cancels the linear phase due to detuning, improves the coherence time, and reduces the shot-to-shot noise.

\subsection{Detection}
Our detection scheme starts with a strong pulse of magnetic gradient to spatially separate the two different spin states. Once separated into two clouds, we apply absorption imaging with an on-resonance laser tuned to the $F=2 \rightarrow F'=3$ transition. To image the atoms in the $F=1$ state, which is a dark state, we send a pulse of the repumper laser before the imaging laser pulse. The shadow of the atoms is captured on a CCD image from which we analyze the optical density profile, fit it to a 2D-Gaussian, and count the number of atoms in each cloud. When required, we apply a low-pass filter to the image to reduce the noise and improve the signal-to-noise ratio.

\subsection{SGI scheme}
We hold one of the two WPs stationary for a duration $T_{\rm hold}$ applying an acceleration equal and opposite to the gravitational acceleration g.
We apply short magnetic gradient pulses with duration $T_{kick}$ to control the second WP to launch it into a ballistic trajectory and stop it at the bottom of the trajectory, so that at the end of this sequence, the two WPs should overlap (up to a small mismatch as shown in Fig.\,\ref{fig:Trajectories_V1.png}) in position and momentum.
Between the pulses, we allow a delay time of $T_d$. In the ideal case, $T_{\rm kick} ,\,T_d$ are very short delta pulses; in an actual experiment, they have a finite length.

Fig.\,\ref{fig:Scheme_V1.png} of the Methods section shows the SGI scheme in terms of RF pulses and magnetic gradients and the trajectories of the WPs. We kick the one WP up, allow it to undergo a ballistic trajectory in free fall, and stop it at the end of the trajectory while we hold the second WP in place. At each moment, we apply a force only on one arm while the other arm is in a state that is first-order magnetically insensitive ($m_F=0$) and unaffected by the magnetic gradients (to first-order).

In this scheme, the force from the atom chip is applied while the atoms are located in the quadrupole of the magnetic field produced by the chip or very close to it. This condition reduces the phase noise and improves the stability of the interferometer.

\subsection{Analytical model for calculating the phase}
In addition to what has been provided in the main text, let us show how the phase difference between the two arms [Eq.\,(4) in the main text] is derived by two methods, one using the Gauge transformation [Eq.\,(1) of the main text], and the second using the action (Secs.\,G.1 and G.2, respectively). In both these derivations the analytical model approximates the experiment as a set of square pulses of a homogeneous magnetic gradient so that the acceleration of each arm is a piecewise constant function. In Sec.\,G.3, we calculate the action of each of the arms from the perspective of both frames, and finally, in Sec.\,G.4, we show how the result may be obtained using the Galilean transformation.

\subsubsection{Derivation of the phase difference using the Gauge transformation}

The transformation between an Einsteinian and Newtonian frames of reference [Eq.\,(1) of the main text] can be generalized to represent a transformation between any frame without forces and a frame with a constant acceleration $\mathbf{a}$, such that
\begin{equation}
 \psi_N(\mathbf{r},t)=e^{i\phi_{\rm gauge}(\mathbf{r},t)}\psi_E(\mathbf{r}-\frac12\mathbf{a}t^2,t), \quad \mathrm{with} \quad \phi_{\rm gauge}({\bf r},t)=\frac{m}{\hbar}[-\frac16a^2t^3+\mathbf{a}\cdot\mathbf{r}]\,.
\label{eq:gentrans} 
\end{equation}
Here it is assumed that both frames have the same velocity at $t=0$ and the zero point for the energy is taken to be at $\mathbf{r}=0$.

First we explain the phase difference in the \textit{Gedanken} experiment [Eq.\,(2) of the main text], from the gauge transformation. We take the time $t=0$ to be at the middle point where the ballistic path is at the turning point. In the frame falling with this path the velocity at this point is equal to the velocity of the Newtonian (laboratory) frame ($v=0$ in the laboratory frame). At the two end-points $t=\pm T$ this frame has a velocity $\mp v_0=\mp gT$ and to fulfil the closing condition the center of the WP at these times should be at the same position of the reference path, which we may choose to be $z=0$. In addition, at these points we apply a kick $v_0 = gT$, which amounts to adding a phase term $e^{\pm imv_0 z/\hbar}=e^{\pm imgTz/\hbar}$ (at $t=-T$ this phase is negative, as the closing condition is reversed in time). This phase cancels the phase in the second term of Eq.\,(1), but it is anyway canceled for the phase at the WP center, which is at $z=0$ at this time.

While Eq.\,(2) of the main text refers to the phase accumulated by the ballistic path during the free-fall time and  represents the full interferometer phase in the \textit{Gedanken} experiment, we can also use the same transformation in Eq.\,(\ref{eq:gentrans}) to derive the total phase difference between the two paths in a model that includes the finite duration of the kick pulses $T_{kick}$, and the delay time $T_d$. In the analytical model that was used in Fig.\,3 we model the gradient pulses in the sequence by square pulses: a kick pulse of duration $T_{kick}$ applied to the ballistic WP, followed by a delay time $T_d$ before the holding pulse of duration $T_h=2T-2T_d$ is applied to the reference WP. The holding time is followed by another delay time $T_d$ and a kick pulse of duration $T_{ kick}$, such that the reference WP is in free fall during the two time intervals of duration $T_{kick}+T_d$ before and after the holding time. We assume that the reference path is completely at rest during the holding time at a constant position $z=0$. It follows that the phase accumulated during the holding time is zero. In order to obtain the phase accumulated during the free-fall time of the reference WP before the holding time we set the time $t=0$ to be just before the beginning of the holding pulse. The gauge phase at this $t=0$ is 0, and the phase comes from the gauge phase at the beginning of the first kick pulse, where the WP is centered at $z_0=-g(T_{kick}+T_d)^2/2$ and we obtain the phase that the reference WP accumulated during the first free-fall interval
\begin{equation}
    \phi_{\rm ref}^{({\rm free})}=-\phi_{\rm gauge}(z_0,-T_{\rm kick}-T_d)=-\frac{m}{\hbar}\left[\frac16g^2(T_{\rm kick}+T_d)^3+gz_0(T_{\rm kick}+T_d)\right]=\frac{mg^2}{3\hbar}(T_{\rm kick}+T_d)^3\,.
\end{equation}

An equal phase is obtained from the  gauge phase during the second free-fall interval of the reference path and we obtain

\begin{equation}
    \phi_{\rm ref}=\frac{2m}{3\hbar}g^2(T_{\rm kick}+T_d)^3\,.
\end{equation}
The phase accumulated by the ballistic WP during the kick pulse can be calculated by using the transformation from a frame that is accelerated by ${\bf a}=(a_{kick}-g)\hat{z}$, where $a_{\rm kick}=g(T-T_d)/T_{kick}$ that satisfies the closing conditions that require that the integrated acceleration applied to the reference WP during the holding time $2(T-T_d)$ should be equal to the integrated acceleration of the ballistic WP. For applying this transformation by using Eq.~\,(\ref{eq:gentrans}) we need to set the initial time $t=0$ to be the time where the velocity is zero in the laboratory frame under the same acceleration. The velocity at the beginning of the first kick pulse is given by $v(t_0)=g(T_{\rm kick}+T_d)$ and therefore the velocity is zero when $t=t_0-v(t_0)/(a_{\rm kick}-g)=t_0-\frac{g}{a_{\rm kick}-g}(T_{\rm kick}+T_d)=0$. At the end of the kick pulse the position of the WP center is at $z_1=z_0+g(T_{\rm kick}+T_d)T_{\rm kick}+\frac12 (a_{\rm kick}-g)T_{\rm kick}^2$ and we obtain for the phase

\begin{equation}
\phi_{\rm kick}=\frac{m}{\hbar}\left\{\frac16(a_{\rm kick}-g)^2 [(t_0+T_{\rm kick})^3-t_0^3]+(a_{\rm kick}-g)[z_1(t_0+T_{\rm kick})-z_0t_0]\right\}\,.
\end{equation}

An equal phase is accumulated during the second kick pulse.
The phase accumulated by the ballistic path during the free fall has to be changed with respect to the expression of the \textit{Gedanken} experiment such that the term linear in time in Eq.\,(\ref{eq:gentrans}) is also included, as in general $z_1\neq 0$. We then have

\begin{equation}
\phi_{\rm ballistic}^{({\rm free})}=2\frac{m}{\hbar}\left[-\frac16 g^2T^3-gz_1T\right]\,.
\end{equation}

When we combine the results of the above equations and substitute for the value of $a_{\rm kick}$ to fulfill the closing condition we finally obtain

\begin{eqnarray} \Delta\phi &=& 2\phi_{\rm kick}+\phi_{\rm ballistic}^{({\rm free})}-\phi_{\rm ref} \nonumber \\
&=& -\frac{mg^2}{3\hbar}\left(T^3+T^2 T_{\rm kick}+T (T_{\rm kick}^2+T_d T_{\rm kick})-T_d(T_{\rm kick}+T_d)^2\right)\,,
\end{eqnarray}
which is  the same result as in Eq.\,(4) of the main text (with an opposite sign due to the fact that the phase in the main text was defined such that it grows positively with the time $T$).

\subsubsection{Derivation of the phase difference using the action}
Alternatively, to calculate the phase of the interferometer in the common method, we use the action of the classical path
\begin{equation}\label{eq:action}
\phi_j = \frac{1}{\hbar}S_j=\frac{1}{\hbar}\int_0^{t_f}\left[K_j(t)-U_j(t)\right] dt ,
\end{equation}
where $j$ is an index of the respective interferometer path, and
\begin{equation}
K_j(t)=\frac{1}{2} m v_j^2(t) \;,\; U_j(t)= - m a_j(t)[z_j(t)-z_0]\, ,
\end{equation}
where $z_0$ is defined by the zero of magnetic potential, $|B(0,0,z_0)|=0$, which is the position of the magnetic field quadrupole in the experiment (not to be confused with the initial position of the atoms $z_i(0)$). In the case of a closed interferometer, the phase does not depend on $z_0$, while for an open interferometer, the phase will depend on $z_0$.
To maintain the symmetry of the trajectories we set the initial conditions of the COM of the WP to be
\begin{equation}
v_z(0) = g(T_{kick}+T_d) \quad ;\quad z(0) = -g(T_{kick}+T_d)^2/2
\end{equation}
We define the total interferometer time as $2T+2T_{kick}$, and the holding duration of the bottom arm can be defined as $T_h = 2T-2T_d$. The experiment can be described by a set of durations $(T_{kick},T_d,T_h,T_d,T_{kick})$, where the acceleration of the ballistic and reference paths in each of these durations is given by $a_{ballistic} = -g +(a_{kick} , 0 , 0 , 0 , a_{kick})$ and $a_{reference} = -g + (0 , 0 , a_{hold} , 0 , 0)$.

The condition for a closed interferometer is an equal area of the accelerations of the two arms,
\begin{equation}
\int_{t_i}^{t_f} a_{1}(t) dt=\int_{t_i}^{t_f} a_{2}(t) d t \,
\end{equation}
which, in this case, gives
\begin{equation}
    a_{kick}\cdot 2T_{kick}= a_{hold}\cdot T_h \,.
\end{equation}
Setting $a_{hold} = g$ confirms that the reference path is stationary in the laboratory frame.
Performing the integrals, we get

\begin{equation}\label{eq:phi1}
\phi_1 =  \frac{m g^{2}}{3 \hbar} \left(- T^{3} - T^{2} T_{kick} - T T_{kick}^{2} - T T_{kick} T_{d} + 2 T_{kick}^{3} + 7 T_{kick}^{2} T_{d} + 8 T_{kick} T_{d}^{2} + 3 T_{d}^{3}\right)
\end{equation}

\begin{equation}\label{eq:phi2}
\phi_2 =  \frac{2m g^{2}}{3 \hbar} \left(T_{kick} + T_{d}\right)^{3}
\end{equation}

\begin{equation}\label{eq:phase_diff2}
\Delta\phi = \frac{mg^2}{3\hbar}\left(T^{3} + T^{2} T_{kick} + T \left(T_{kick}^{2} + T_{kick} T_{d}\right) - T_{d} \left(T_{kick} + T_{d}\right)^{2}\right).
\end{equation}

Taking the limit of a short kick we get
\begin{equation}\label{eq:phase_diff_lim_T1}
\lim_{T_{kick}\to 0} \Delta\phi = \frac{mg^2}{3\hbar}\left(T^{3} - T_{d}^{3}\right),
\end{equation}
and the limit of short delay we get
\begin{equation}\label{eq:phase_diff_lim_Td}
\lim_{T_{kick},T_d\to 0} \Delta\phi = \frac{mg^2}{3\hbar}T^{3}.
\end{equation}
Eq.\ref{eq:phase_diff_lim_Td} represents the \textit{Gedanken} experiment in which the beam splitter is a delta pulse such that  $T_{kick},T_d\to 0$.

The result of the analytical model given in Eq.\,(\ref{eq:phase_diff2}) (and in Eq.\,(4) of the main text) is plotted in Fig.\,3 of the main text using the experimental values of $T_{kick} = 80\rm\,\mu s$ and $T_d = 71+6\rm\,\mu s$ where we include half the rise time of the holding pulse $\tau_{\rm hold}/2 = 6\rm\,\mu s$ in the delay. We also include the contribution to the phase from the force on the state $|0\rangle$ due to the second-order Zeeman (SOZ) effect. This force is opposite to the force on the $|1\rangle$ state during the gradient pulses, and its value is about 1\% of the latter. During the holding pulse, the acceleration of the state $|0\rangle$ due to SOZ is calculated to be ${a}_{SOZ}=0.10\,\rm m/s^2$ (see Sec.\,K) for details). We include this effect in the analytical prediction for the phase (Eq.\,17) by replacing the gravitational acceleration $g$ by $g_{\rm eff}=g+a_{\rm SOZ}=9.91\,\rm m/s^2$. This estimation agrees with a model in which $a_{SOZ}$ is explicitly included in the calculation to within $0.03\,\rm rad$.

An analytical calculation can also include the temporal shape of the pulses and account for time-dependent homogeneous magnetic pulses. However, for the scope of this paper, we choose to show the analytical case only for the more straightforward case of piecewise constant acceleration and use the numerical simulation to include the temporal and spatial modulation of the gradients.

\subsubsection{Calculating the actions in both frames}

Next, to understand the symmetry of the QGI, as emphasized in the introduction, we calculate for the {\it Gedanken} experiment the actions associated with the classical trajectories of the reference and the ballistic WPs as seen from both frames of reference, keeping track of the contributions from the kinetic and potential energies as well as the pulses, which we model for the sake of simplicity by delta functions.\\
\indent
We start by discussing the situation in the Newtonian frame. Here the reference WP remains at rest for all times, since the gravitational acceleration $g$ is compensated by the acceleration $a$ provided by the magnetic field gradient. Hence, the contributions due to the kinetic and potential energies vanish. Moreover, there are no pulses. Thus, the total action of the reference WP vanishes as indicated in Table~\ref{tab:2}.\\
\indent
In contrast, two parts contribute to the total action of the ballistic WP. Due to the non-vanishing momentum $m v_0$ of the atom originating from the first pulse, the accelerated motion in the linear gravitational potential provides us with non-vanishing actions due to the kinetic as well as the potential energies. Their structure shown in the second row of Table~\ref{tab:2} becomes apparent when we recall that the velocity of this motion consists of the constant velocity $v_0$ and the one caused by the gravitational acceleration $g$. The action due to this kinetic energy involves the integral of the square of this time-dependent velocity creating the sum of three terms:
($i$) The kinetic energy $m v_0^2/2$ multiplied by the interferometer time $2T$. ($ii$) A cross term between the velocity $v_0$ and the acceleration $g$ and ($iii$) the square of the acceleration times  $(2T)^3$.\\
\indent
Moreover, we emphasize that each term has an interesting combination of $v_0$, $g$ and $T$. Indeed, the term proportional to $v_0^2$ is linear in $T$ but independent of $g$. The cross term is linear in $v_0$ and $g$ but quadratic in $T$, and the third contribution is independent of $v_0$ but quadratic in $g$ and cubic in $T$.\\
\indent
This scaling in $v_0 g$ and $g^2$ is crucial. Indeed, when we replace $v_0$ by $-v_0$ and $g$ by $-a$ as defined by the Einsteinian frame discussed in the second column of Table~\ref{tab:2}, the expressions, now occurring for the reference rather than the ballistic WP, are identical.\\
\indent
Next we turn to the contribution from the potential energy of the ballistic WP in the Newtonian frame. Since the ballistic WP starts from the origin of the coordinate axis, the motion is a combination of the linear motion due to $v_0$, and the gravitational acceleration $g$ giving rise to two terms in the action which are identical in size and sign to the last two terms of the action due to the kinetic energy. Hence, these terms also satisfy the symmetry in $v_0g$ and $g^2$.\\
\indent
It is interesting that in the Newtonian frame the delta function pulses do not contribute to the actions of the ballistic WP in the Newtonian frame, and this for two reasons: First, the corresponding potential energy is proportional to the coordinate at the time of the pulse and second, the ballistic WP starts at $z = 0$ and concludes its journey at $z = 0$. \\
\indent
As a result, the total action of the ballistic WP is given by the ones of the kinetic and potential energies which due to the closing condition connecting $v_0$ and $T$ allows us to combine the three terms in a single one shown in the Table~\ref{tab:2}, determining the difference of the actions corresponding to the two WPs.\\
\indent
Finally, we address the Einsteinian frame which is attached to the motion of the ballistic WP. As mentioned before, the two WPs exchange their roles compared to the Newtonian frame. Hence, now the action of the ballistic WP vanishes but the one of the reference WP is non-vanishing arising from the magnetic field gradient in the same way as for the ballistic WP in the Newtonian frame.\\
\indent
Moreover, the symmetry of the terms discussed before, ensures that the origin of this contribution is identical to one of the terms of the ballistic WP in the Newtonian frame.\\
\indent
Since we consider the difference of the actions of the two WPs there must be a sign difference between the two frames. Indeed, in the Newtonian frame the action of the reference WP vanishes, whereas for the Einsteinian frame it is that of the ballistic WP.

\subsubsection{Derivation of the transformation phase between the Newtonian and Einsteinian frames using Galilean transformation}

Here we re-derive the transformation phase between the Newtonian and Einsteinian frames using the Galilean transformation without relying in any way on the EP. For this purpose, we do not assume equality of the inertial and gravitational masses and use the Newton equation $m_i\ddot{z}_{cl}=-m_g g$, to connect 
\begin{equation}
 \label{G4_1}
    \psi_N(z,t)=e^{i\chi(z,t)}\psi_E\left(z-v_0 t+\frac{1}{2}\frac{m_g}{m_i}gt^2,t\right),\; \text{where}\;\;\;\chi(z,t)\equiv \chi_0(t)+\chi_1(t)z
\end{equation}
the time-dependent wave function $\psi_N$ in the Newtonian frame and the one $\psi_E$ in the Einsteinian frame, that is the one moving along the classical trajectory $z_{cl}(t)=v_0 t-(1/2)(m_g/m_i)gt^2$ with a initial velocity $v_0$. For writing the phase $\chi(x,t)$ as the linear function of $z$ with two unknown functions of time $\chi_0(t)$ and $\chi_1(t)$, we have used the fact that the particle always increases its momentum in a linear potential. 

To derive the analytical forms of $\chi_0(t)$ and $\chi_1(t)$, one has to insert the function $\psi_N(z,t)$ into the Schr\"odinger equation 
\begin{equation}
 \label{G4_2}
    i\hbar\frac{\partial}{\partial t}\psi_N(z,t)=\left(-\frac{\hbar^2}{2m_i}\frac{\partial^2}{\partial z^2}+m_g gz\right)\psi_N(z,t)
\end{equation}
in the Newtonian frame and use then the Schr\"odinger equation
\begin{equation}
 \label{G4_3}
    i\hbar\frac{\partial}{\partial t}\psi_E(z',t)=-\frac{\hbar^2}{2m_i}\frac{\partial^2}{\partial z'^2}\psi_E(z',t)
\end{equation}
in the Einsteinian frame, the particle is at rest in this frame but undergoes free quantum dynamics.

As a result, we arrive at 
\begin{equation}
 \label{G4_4}
    \begin{split}
        \chi_0(t) =& -\frac{1}{\hbar}\int_{0}^{t}d\tau \frac{\left(m_i v_0-m_g g\tau\right)^2}{2m_i}=-\frac{1}{2\hbar m_i}\left(m_i^2 v_0^2 t-m_i m_g v_0 gt^2+\frac{1}{3}m_g^2 g^2 t^3\right)\\
         \chi_1(t)=&\frac{m_i v_0-m_g gt}{\hbar}.
    \end{split}
\end{equation}

In the case of $v_0=0$ and $m_i=m_g$, Eq. \eqref{G4_1} gives Eq.\,(1) of the main text.

Finally, we obtain the phase difference measured by the QGI given by Eq.\,(2) of the main text, using only the Galilean transformation, Eq. \eqref{G4_1}. To do it, we consider the wave function $\psi_N$ at $t=2T$ and use the closing condition $T=m_iv_0/(m_g g)$. As a result, the spatial argument in $\psi_E$ reduces to $z$, corresponding to perfect overlap of the reference and ballistic WPs in the coordinate space. Moreover, the second pulse employed at $t=2T$ imprints again the momentum $p_0=m_i v_0$ onto the wave function and thus cancels the position-dependent phase $\chi_1(2T)z=-m_iv_0 z/\hbar$. Hence, the QGI is closed in both momentum and position space. The phase difference therefore coincides with the phase between $\psi_N(z,2T)$ and $\psi_E(z,2T)$, that is
\begin{equation}
\psi_N(z,2T)=e^{i\phi_{ballistic}}\psi_E\left(z,2T\right),
\end{equation}
where the ballistic phase is entirely determined by $\chi_0(2T)$ and reads
\begin{equation}
    \phi_{ballistic}=\chi_0(2T)=-\frac{1}{2\hbar m_i}\left(\frac{8}{3}m_g^2 g^2 T^3-2m_i^2 v_0^2 T\right)=-\frac{1}{2\hbar m_i}\left(\frac{8}{3}-2\right)m_g^2 g_2 T^3=-\frac{m_g^2}{3\hbar m_i}g^2T^3.
\end{equation}
Here we have used the closing condition, $T=m_iv_0/(m_g g)$.

\begin{center}
\begin{table}[H]
\caption{\textbf{Contributions to the actions of the center-of-mass motion of the reference and the ballistic WPs in the QGI viewed from the Newtonian and the Einsteinian frames.} Note that the closing conditions to obtain the respective total actions are written in round brackets whereas the levitation condition for the respective difference in actions are given by the expression in square brackets. In the Newtonian frame the reference WP is levitated and remains at rest at the origin. The Einsteinian frame is attached to the ballistic WP. Thus, in this frame the WP remains at rest at the origin yielding only vanishing actions. This is also the case for the reference WP but in the Newtonian frame.}

\footnotesize

\renewcommand{\arraystretch}{2.5}
\centering
\begin{tabular}{ |c  c|c |c| c c| }

\hline
\multicolumn{2}{|l|}{\diagbox[width = 4cm]{actions}{frame}}&\makecell{Newtonian \\ \vphantom{($g = 0$)}} &  \makecell{Einsteinian \\ \vphantom{($g = 0$)}} \\
\hline
\multirow{4}{4em}{\makecell{reference \\ WP}}
& kinetic
& 0
&$ m v_0^2 T - 2m v_0 a T^2 + \frac{4}{3} m a^2 T^3$\\
&potential
&0
& $- 2 m v_0 a T^2 + \frac{4}{3} m a^2 T^3$\\
&pulses
&no pulses
&0\\
&\makecell{total\\ \vphantom{$(v_0 =  a T)$}}
&\makecell{0\\ \vphantom{$(v_0 = a T)$}}
&\makecell{$-\frac{1}{3} m a^2 T^3$\\ $(v_0 =  a T)$}
\\[4mm]
\hline

\multirow{4}{4em}{\makecell{ballistic \\ WP}}
& kinetic
& $ m v_0^2 T - 2 m v_0 g T^2 + \frac
{4}{3} m g^2 T^3$
&0
\\
& potential
& $-2 m v_0 g T^2 + \frac{4}{3} m g^2 T^3$
& 0  \\
& pulses
& 0 & no pulses
\\
&\makecell{total\\ \vphantom{$(v_0 = \frac{1}{2} a T)$}}
& \makecell{$-\frac{1}{3}m g^2 T^3$\\[2mm]($v_0 = g T$)}
& \makecell{0\\ \vphantom{$(v_0 = \frac{1}{2} a T)$}}
\\[4mm]
\hline
\multicolumn{2}{|l|}{\makecell{difference in\\ total actions}}
& \makecell{$ -\frac{1}{3}m g^2 T^3$\\[1mm] \vphantom{$[m a = - m g]$}}
& \makecell{$\frac{1}{3} m a^2 T^3$ \\[1mm] $[a = - g]$}
\\[4mm]
\hline
\end{tabular}
\label{tab:2}
\end{table}

\end{center}

\subsection{Numerical simulation}\label{sec:numerical_simulation}

\subsubsection{Details of the numerical simulation}

The numerical simulation calculates the dynamics of the atomic BEC during the trapping, release, DKC, and interferometer sequence. The simulation is based on the model of WP evolution given in\,\cite{japha2021}, which is extended to include the effects of rotations. To use the model with curved magnetic potentials, we locally approximate the magnetic potential by a quadratic potential and calculate the evolution of each WP in the potential. The simulation predicts the output of the SGI by calculating the overlap integral of the two final WPs, resulting in a complex value, where the visibility is the magnitude of this integral and the phase is the argument.

The simulation is constructed to mimic the exact experimental sequence by using the same timings as in the experiment. During the preparation stage, the currents and magnetic fields in the simulation are kept close to the values reported in the experiment but fine-tuned to match the WP's initial conditions, such as initial position and velocity, and initial expansion rate. As the measurements of the absolute positions of the atoms are accurate up to a few $\rm \mu m$, other indications of the position of the atoms in the lower interferometer arm are also taken from the holding current required to hold the atoms against gravity, and the trajectory after the DKC. A holding current of 23-24\,mA corresponds, according to the calculation, to a distance of 112-113\,$\mu$m from the chip surface. Assuming that the velocity of the atoms at this position before turning on the holding pulse is almost zero, we can deduce the trajectory after the DKC as the atoms are in free fall and hence travel a distance $\Delta z=\frac12 g t^2$. This calculation indicates a distance of about 132\,$\mu$m from the chip surface for the initial trapping position. These values of the initial trap and turning point agree with direct measurements from the experiment.

The magnetic bias field is taken to be the field generated by the bias coils and Earth's magnetic field ${\bf B}_E=(-0.31,0,0.33)$\,G. The bias generated by the coils consists of a component in the $y$ direction and in the $x$ direction, which are determined to be $B_{0y}=12.9928$\,G and $B_{0x}=-1.175$\,G. The $y$-component of the field during the trapping determines the trap position with respect to the chip surface and the $x$-component determines the trap frequency and field at the bottom of the trap. The total magnetic field magnitude is indeed measured some time after trap release by the RF transition frequency between the Zeeman levels $|2,2\rangle$. and $|2,1\rangle$. This measurement indicates that the total field at this time is $B_0=12.625$\,G, but we take the field during the trapping to be $B_0=13.05$\,G to ensure that the trapping position is about 132\,$\mu$m from the chip surface, which is consistent with the position of the atoms in the lower arm of the interferometer, as described above. The discrepancy between the measured bias field after the trap release and the predicted bias field required to explain the trapping stage may be attributed to the magnetic response of the atom chip mount and the chamber itself, resulting in transient fields in the science chamber during the trapping. Time dependence of the bias field after the trap release and DKC is observed in the RF resonance measurements.
The $z$-component of the magnetic field generated by Earth is responsible for a shift of the trapping position relative to the center of the system and we find numerically that the trapping position is at ${\bf r}_{\rm trap}=(0.79,4.67,132.05)\,\mu$m relative to the center of the surface of the central wire. The trapping frequencies are $\omega_y\approx \omega_z\approx 2\pi\times 1.04$\,kHz and $\omega_x\approx 2\pi\times 57$\,Hz, which is in good agreement with the measured trap frequencies. The calculated field at the bottom of the trap is $B_0\approx 0.825$\,G, which is slightly larger than the measured value. For a total number of $2\cdot 10^4$ atoms, we obtain a BEC of an average width $\sigma_x=3.13\,\mu$m in the longitudinal direction and $\sigma_y=\sigma_z=1.31\,\mu$m in the transverse direction. This serves as the starting point for calculating the evolution of the WP widths during the sequence. The constant bias field also includes a measured gradient $B'=68.32$\,G/m, giving rise to an acceleration $\mu_B B'/2m=0.22$\,m/s$^2$ in the direction of gravity.
The DKC pulse kicks the atoms upwards, and after it, the atoms are in free fall until the beginning of the interferometer, except that they are affected by small magnetic gradients from the bias field and an idle current from the kick wires on the chip. During this time, they are transferred from the state $|2,2\rangle$ to the state $|2,1\rangle$ by a RF pulse that ends about 400\,$\mu$s after the collimation pulse and then they are split into a superposition of $|2,1\rangle$ and $|1,0\rangle$. As the idle current is designed to counteract the magnetic gradient of the bias during the time between the splitting $\pi/2$ pulse and the recombining $\pi/2$ pulse we may consider the trajectory of the atoms before the kick pulse as being ballistic and calculate the timing and position of the turning point where the atoms are expected to have a zero velocity. This point $z_h$ is located at a distance $z_h-z_{\rm DKC}=\frac{1}{2}gt_{\rm free}^2$ from the position $z_{\rm DKC}$ at the end of the collimation pulse. At this point, the holding current $I_h$ is calculated and the kick current $I_k$ is also calculated such that the holding current and kick currents satisfy $2I_kT_k=I_h T_h$. The DKC is optimized to yield a zero velocity of the atoms at the beginning of the holding pulse such that the lower trajectory is constant during the holding time. The ballistic trajectory is calculated based on the calculated kick current and the second-order Zeeman effect is taken into account during the holding time of the lower arm, as well as the effect of the second-order Zeeman over the whole sequence.

The free parameters of the simulation are the bias field in the x and y directions and the exact value of the collimation current. When these parameters are varied under the constraint that the lower arm trajectory is flat and at the correct distance from the chip surface, the interferometer phase varies within a range of $\pm 0.07$\,rad at the longest holding time. Note that these parameters were not used for fitting the phase but rather to fit other parameters of the experiment mentioned above; thus the numerical simulation curve in Fig.\,3 can be considered as a curve without fitting parameters related to the phase.

In addition, the second simulation curve [blue line in Fig.\,3(B)] represents an attempt to best fit the experimentally measured phase without exceeding the uncertainty range of the parameters, with the price of the lower arm trajectory not being flat. In the latter simulation, the holding current was taken to be 23.5\,mA (although a larger current was required to hold the reference arm against gravity), and the DKC current was taken to be 0.4705\,mA (instead of 0.4695\,mA for the first simulation). The kick current was taken larger by 0.15\%, the idle current was taken to be 0.5\,mA instead of 0.55\,mA, and $B_z$ was taken to be 0.35\,G instead of 0.33\,G. All these changes are well within the uncertainty limits of the experimental parameters.

\subsubsection{Effect of WP shape on the phase}
\begin{figure}[H]
\centering
\includegraphics[width=1\textwidth]{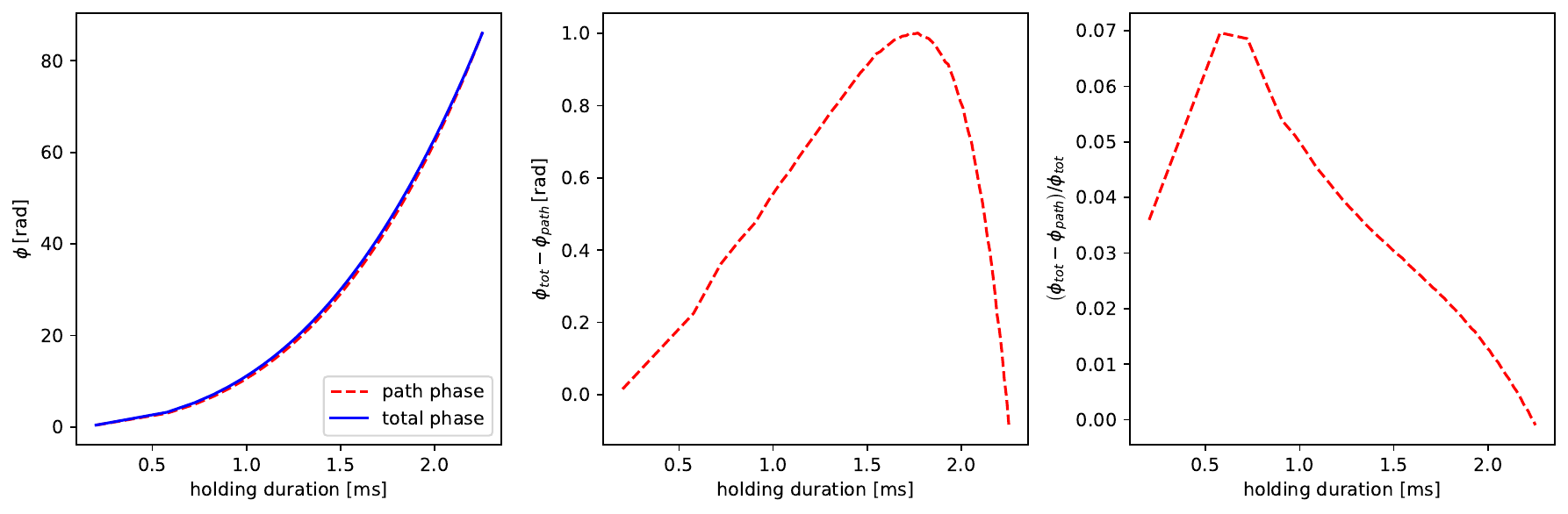}
\caption[]{The different phase contributions calculated in the numerical simulation. (A) The total phase (blue solid line), was calculated from the overlap integral between the two WPs at the end of the interferometer (the timing of the last $\pi / 2$ pulse). This total phase takes into account effects of WP shape and atom-atom interaction. The red dashed line is the ``path phase", the phase accumulation due to the classical trajectory of the COM. (B) The difference between the two calculations, shows that the ``non trajectory" contributions range between $-1.1$ to $0.4 \rm\,rad$, out of the total $80\rm\,rad$ of the interferometer. (C) The relative part of the ``non trajectory" is as high as $15\%$ for short interferometer duration, but reduces to less than $2\%$ for durations longer than $1.2\rm\,ms$.}
\label{fig:Numerical_simulation_phase_trems.pdf}
\end{figure}

The numerical simulation includes the effect of the wave-packet shape, which evolves due to single particle effects (internal kinetic energy at the wave-packet center and response to the curved potentials) and atom-atom interactions. To estimate the effects of the wave-packet shape on the phase, we compare the total phase calculated from the overlap integral to the phase of the classical trajectory of the COM. The result is shown in Fig.\,\ref{fig:Numerical_simulation_phase_trems.pdf}. The total phase accumulation in the experiment amounts to $80\rm\,rad$ while the contribution due to the shape of the WP is in the range of $0$ to $1 \rm\,rad$. The relative contribution of the ``non trajectory" phase is as high as $7\%$ for short interferometer durations, but reduces for longer interferometer times. The exact contribution due to the shape of the WP depends on the exact shape of the WP and the specific experimental conditions, and can be one of the sources of the discrepancy between the numerical simulation and the experimental data.
The numerical simulation also includes the effect of the ambient gradient in the chamber, which was measured to be $0.7\,\rm G/cm$ (equivalent to an acceleration of $0.2\,\rm m/s^2$ along gravity of the $|1\rangle$ state, see Sec.\,L for more details) and the idle current of $0.55\,\rm mA$ we apply to counteract the effect of the ambient gradient.
The shape of the pulses of current in the SG wires is also included in the simulation. A detailed description of the pulses is given in Sec.\,J.

\subsection{Data analysis for figures 2 and 3}
To extract the phase from the raw data of population vs. time, we fit the data to a model $P = P_{\text{mean}}(2T) + \frac{1}{2}V(2T)\cos\left[\phi(2T)\right]$, where $V$ is the visibility of the oscillations, $\phi$ is the phase of the oscillations and $P_{\text{mean}}$ is the mean value of the population. To obtain the fit, we first identify the upper and lower envelopes and fit them to a polynomial to get $P_{\text{mean}}$ and $V(2T)$. We then fit the data to the model where the phase $\phi(2T)$ is a third-order polynomial. During the fitting we exclude the first and last oscillation as they suffer from larger uncertainty in the values of the envelope.
As fitting such a model to the data requires a good initial guess, we use the Hilbert transform \cite{Whittaker2015, Rozenman2021} to extract the phase from the data. The details of the analysis are given below.


\subsubsection{Figure 2 - Extracting the phase and fitting the data}
We fit the data of population vs. time with the following algorithm: First, we fit the envelope of the oscillations by finding the local maxima (minima), connecting them with a set of straight lines, and fitting them to a polynomial. For the data in Fig.\,2, a polynomial of order seven was used for the envelopes. Once we have an equation for the upper and lower envelope, we calculate the local amplitude $V/2(2T)$ and local mean $P_{mean}(2T)$. We then apply the Hilbert transform, which results in a complex number for each data point, whose argument is the phase up to $2\pi n$, so we unwrap the phase of the whole data, which requires high enough sampling rate in each oscillation. At this point, we fit a third-order polynomial to obtain a smooth function $\phi(2T)$. We use this polynom as the initial guess for the final fit on the population. While the Hilbert transform is a powerful tool for extracting the phase, it introduces significant edge effects, which are hard to avoid. The direct fit of the phase to the data is more robust to edge effects, and is used for the final fit of the data.

\subsubsection{Figures 3}
In Fig.\,3(A), we show the phase derived from the experimental data against three calculations of $\phi(2T)$: the numerical simulation, the analytical approximation and the expected phase in the {\it Gedanken} experiment. We calculate the lines of the analytical approximation and the \textit{Gedanken} experiment using Eq.\,\ref{eq:phase_diff} and Eq.\,\ref{eq:phase_diff_lim_Td} respectively. The line of experimental data was derived by a fit to the data of Fig.\,2 (as explained above), and the data points of the numerical simulation were derived as explained in Sec.\,H. We also plot as a guide to the eye deviations from the $T^3$ phase accumulation, by plotting a modified version of the analytical approximation,
\begin{equation}
\Delta\phi =  \frac{mg^2}{3\hbar}\left[\left(\frac{mg^2}{3\hbar}\right)^{\frac{\alpha}{3}}T^{3+\alpha} +T^{2} T_{kick} + T \left(T_{kick}^{2} + T_{kick} T_{d}\right) - T_{d} \left(T_{kick} + T_{d}\right)^{2}\right],
\end{equation}
where we set $\alpha = 0.15$ to show 5\% deviation from the $T^3$ phase accumulation, and the coefficient $\frac{mg^2}{3\hbar}^{\frac{\alpha}{3}}$ is added to maintain the correct units.
The conditions that define the phase in interest are that the interferometer is closed and the reference arm is stationary. The closing condition is best approximated by the condition of equal areas of current (Eq.\,36). We found experimentally, analytically, and numerically, that phase is much more sensitive to small deviations from the equal areas of current than to small deviations from the flatness of the reference arm. Thus to estimate the uncertainty bounds of the phase, we repeat the experiment with an increase (decrease) of the magnetic-pulse current $I_{kick}$ by $0.5\,\%$ (which is our estimated experimental precision on this parameter). We use the result to plot the upper (lower) uncertainty bounds of the phase.
The inset shows the rate of phase accumulation $\frac{\partial \phi}{\partial t}$, of the same four cases. To get the derivative of the phase for the numerical simulation, we fit a third-order polynomial on the set points and calculate the analytical derivative of the polynomial. For the experimental data we use the analytical derivative of the polynomial found in Fig.\,2. The derivative of the analytical approximation and the \textit{Gedanken} experiment is calculated by the derivative of Eq.\,\ref{eq:phase_diff} and Eq.\,\ref{eq:phase_diff_lim_Td} respectively. We also plot in the inset the 5\% deviation from the $T^3$ phase accumulation, this time by plotting the derivative of the modified \textit{Gedanken} experiment phase accumulation.
\begin{equation}\label{eq:dphi_dt_with_alpha}
    \Delta\phi(2T) = \left(\frac{mg^2}{24\hbar}\right)^{\frac{3+\alpha}{3}}(2T)^{3+\alpha} \rightarrow  \frac{\partial \phi}{\partial (2T)} = \left(3+\alpha\right)\left(\frac{mg^2}{24\hbar}\right)^{\frac{3+\alpha}{3}}(2T)^{2+\alpha},
\end{equation}
where the power of the coefficient is corrected to maintain the correct units.
Fig.\,3(B) shows the residuals between the experimental data and the numerical simulation. To estimate the statistical noise of the phase, we use a method of error propagation from the standard deviation (STD) of the population data to the STD of the phase, which is explained in the next section.

Finally, the analysis shows that the predicted prefactor of $-\frac{1}{3\hbar}mg^2$ is consistent with the data.

\subsubsection{Error analysis}\label{sec:error_analysis}
We wish to find the phase uncertainty due to the uncertainty of a cosine signal $S(\phi)=a \cos (\phi)$, where the amplitude $a$ and the phase $\phi$ have an uncertainty $\delta a$ and $\delta \phi$, respectively. The uncertainty of the signal due to the phase and amplitude uncertainties is given by

\begin{equation}
\delta S^{2}=\delta a^{2} \cos ^{2} \phi+a^{2} \sin ^{2} \phi \delta \phi^{2}.
\end{equation}

This is a familiar relation, but it is important to remember that the trigonometric functions on the right-hand side should be averaged over the uncertainty $\delta \phi$ of the phase (otherwise this equation is not correct and gives wrong values, especially around the extremum points of the signal, where the derivative approaches zero). Assuming that $\phi$ is a Gaussian distributed variable with a standard deviation $\delta \phi$, the expectation values of the trigonometric functions are

\begin{equation}
\left\langle\sin ^{2} \phi\right\rangle=\frac{1}{2}[1-\langle\cos (2 \phi)\rangle], \quad\left\langle\cos ^{2} \phi\right\rangle=\frac{1}{2}[1+\langle\cos (2 \phi)\rangle] .
\end{equation}

where

\begin{equation}
\langle\cos (2 \phi)\rangle_{\phi=\phi_{0}}=\frac{1}{\sqrt{2 \pi} \delta \phi} \int_{-\infty}^{\infty} d \phi e^{-\left(\phi-\phi_{0}\right)^{2} / 2 \delta \phi^{2}} \cos (2 \phi)=\cos \left(2 \phi_{0}\right) e^{-2 \delta \phi^{2}}
\end{equation}

We obtain

\begin{equation}
\delta S^{2}=\frac{1}{2}\left[1+\cos (2 \phi) e^{-2 \delta \phi^{2}}\right] \delta a^{2}+\frac{1}{2} a^{2}\left[1-\cos (2 \phi) e^{-2 \delta \phi^{2}}\right] \delta \phi^{2}.
\end{equation}

\begin{figure}[H]
    \centering
    \includegraphics[width=1\textwidth]{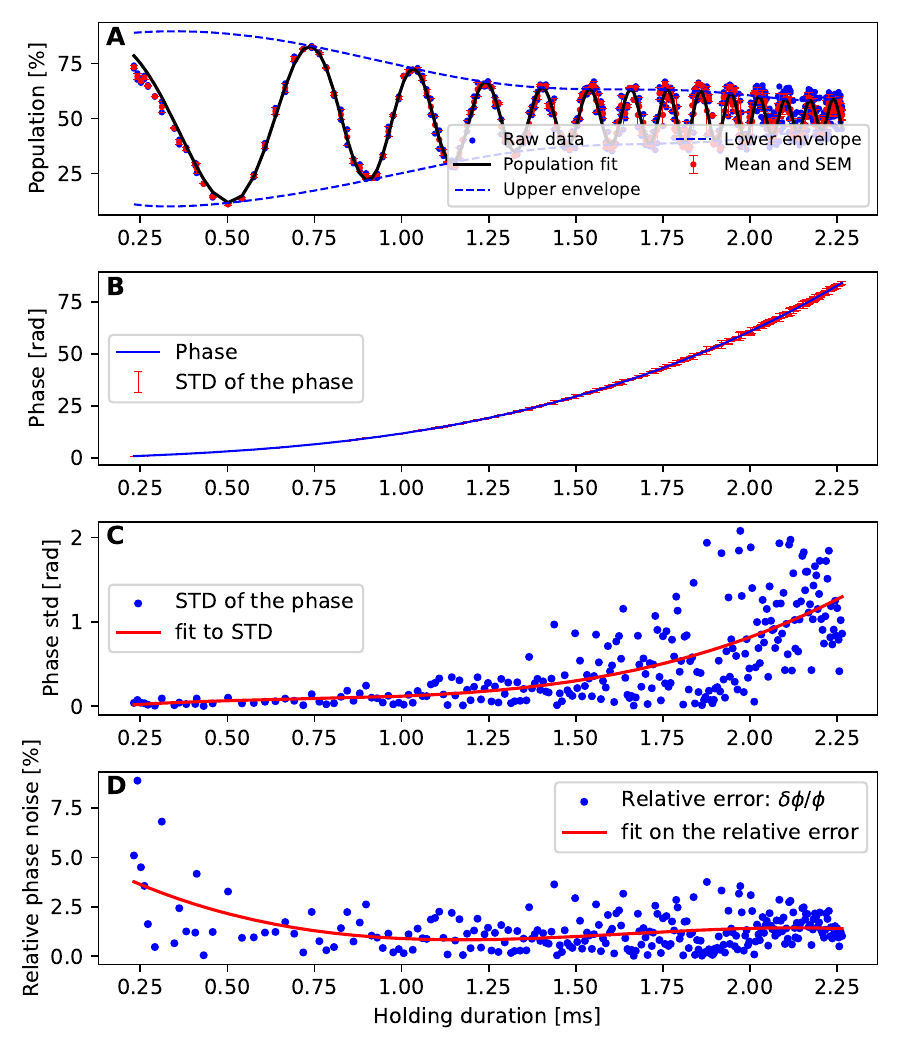}
    \caption[]{Summary of the error analysis. (A) The experimental data, including the raw data of the population vs. time, the fit to the envelope, and the fit to the data.
    (B) The phase as a function of the holding duration, with error bars indicating the STD of the phase. The STD is calculated using the method of error propagation explained in Sec.\,I3. (C) The STD of the phase, and fit to a third order polynomial. The result of the fit is used as the STD of the phase in Fig.\,3B (D) The relative error of the phase, calculated as the ratio of the STD of the phase to the phase, which has an average value of $1.3\%$.}
    \label{fig:data_analysis_summary}
\end{figure}

To extract the STD of the phase, we need to know the STD of the signal ($\delta S^{2}$), the amplitude and phase of the signal ($a$ and $\phi$), which we get from the fit to the data, and the STD of the amplitude ($\delta a$). If we want to find both amplitude and phase uncertainties from the signal uncertainties for a single point this seems to be impossible because at the right-hand-side we have two variables. So we must either assume a given value of $\delta a$ and then find $\delta \phi$ from the signal uncertainty or we should assume some relation of $\delta a$ and $\delta \phi$ along the different measurement points that may enable determination of both uncertainties from the signal. In our case, we assume that the amplitude uncertainty is negligible compared to the phase uncertainty, so we can find the phase uncertainty from the signal uncertainty. This assumption holds well for $2T>1ms$, where the phase noise is the dominant source of noise in the signal, and is approximately proportional to the phase itself. While for $2T<1\,\rm ms$ this assumption is less accurate, the STD of the signal is very small in this range anyway. As the number of iterations per point is 2-4, resulting in large fluctuations of the STD, we fit the STD to a polynomial of order 3, and use the result of the fit as the STD of the phase in Fig.\,3(B). The results of the error analysis are shown in Fig.\,\ref{fig:data_analysis_summary}.

\subsection{Magnetic gradient pulses}\label{section:Magnetic gradient pulses}
The pulses of the magnetic gradient are formed by running a current through three parallel wires, with alternating current orientation, forming a magnetic quadrupole capable of producing high magnetic gradients with low magnetic fields.
The wires are $2\rm\,\mu m$ thick, $40\rm\,\mu m$ wide, and are positioned $100\rm\,\mu m$ apart (center to center), creating a quadrupole $97\rm\,\mu m$ below the surface of the chip (see Fig.\,\ref{fig:BGU2 chip wires layout.png}).
The currents are produced by a homemade current source, with high stability and fast response, controlled by an analog signal input. The current source has a maximum slew rate of $30\rm\,{mA / \mu s}$, a maximum current of $1\rm\,A$ and stability of $100\rm\,ppm$.
We control the current source by an arbitrary waveform generator, which allows us to create smooth and accurate pulses.
To obtain a smooth pulse, we demand that the time derivative $\partial I/\partial t$ is continuous and vanishes at the beginning and end of the pulse.
We choose to use a cosine pulse shape, as it fulfills the conditions of a smooth pulse and is analytically solvable when calculating trajectories and phases.
The equation describing the current for a pulse starting at $t_0$, rise and fall times of $\tau$, separated by a duration $w$ of constant current $A$ is given by
\begin{equation}\label{eq:I_kick}
  \mathbf{I}(\mathbf{t}): \begin{cases}0 & \text { for } t<t_0 \\ \frac{I_{\rm max}}{2}\left(1-\cos \left(\frac{\pi\left(t-t_0\right)}{\tau}\right)\right) & \text { for } t_0<t \leq t_0+\tau \\ I_{\rm max} & \text { for } t_0+\tau<t \leq t_0+\tau+w \\ \frac{I_{\rm max}}{2}\left(1-\cos \left(\frac{\pi\left(t-t_0-w\right)}{\tau}\right)\right) & \text { for } t_0+\tau+w<t<t_0+2 \tau+w \\ 0 & \text { for } t \geq t_0+2 \tau+w\end{cases}
\end{equation}
The area under such a pulse is $I_{\rm max}(\tau+w)$. The scheme has two short kick pulses, and one longer holding pulse. The kick pulses have a rise and fall time of $\tau_{\rm kick}= 40\rm\,\mu s$ each, with $w=0$ and $I_{\rm max} = I_{\rm kick}$, so that the area of the pulse is $I_{\rm kick}\cdot \tau_{\rm kick}$. The longer holding pulse has a rise and fall time of $\tau_{\rm hold} = 12\rm\,\mu s$, and a duration $w=T_h - \tau_{\rm hold}$ of constant current $I_{\rm hold}$ so that the total area of the pulse is $I_{\rm hold}\cdot T_h$.

To maintain the symmetry of the scheme and satisfy the closing condition, the momentum given by the two kick pulses should be equal to the momentum given by the holding pulse, which gives the condition
\begin{equation}\label{eq:a_kick}
    a_{kick} = \frac{ a_{\rm hold} }{2 \tau_{\rm kick}}T_{h}
\end{equation}

\begin{figure}[h!]
    \centering
    \includegraphics[width=1\textwidth]{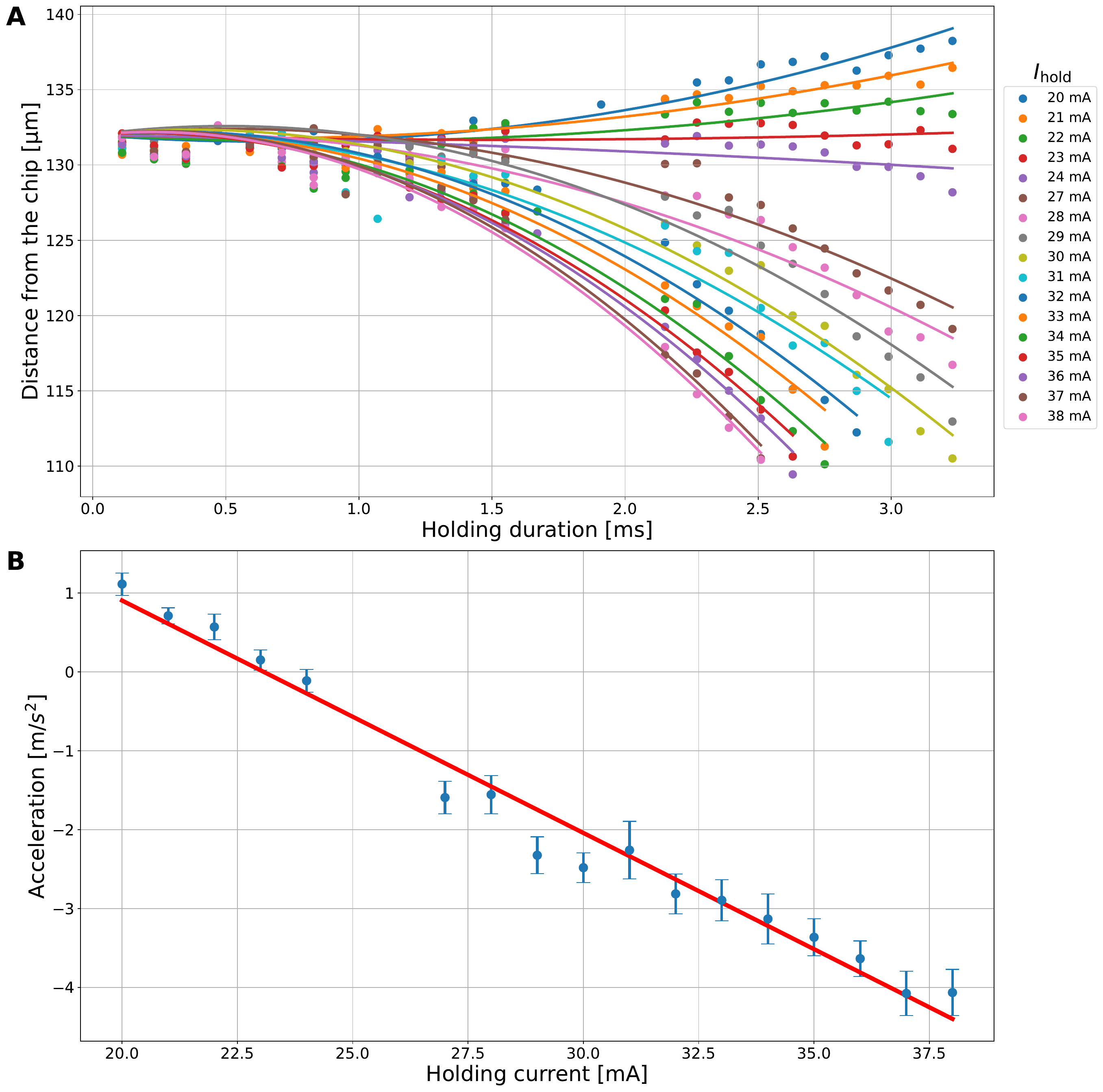}
    \caption[Calibration of the holding current]{Calibration of the holding current: (a) Position along the z-axis vs. time for different holding current values. Each data set is fitted with a parabola of the form $z(t) = z_0+v_0 t + \frac{1}{2}at^2$, where we set $z_0 = 132$ for all the data sets (note that the chip's position in this measurement was not calibrated to $z=0$. As we are interested in the shape of trajectories, the constant shift in z is irrelevant). (b) The acceleration of the atoms vs. the holding current. A linear fit predicts a zero acceleration for $I_{\rm hold} = 23\,\rm\pm 1.5 mA$.}
    \label{fig:holding_current_calibration}
\end{figure}

The acceleration during the holding pulse $a_{\rm hold}$ is tuned to hold the $|1\rangle$ state against gravity (the levitation condition). To calibrate the holding current $I_{\rm hold}$, so that it satisfies the levitation condition, we measured the acceleration of the atoms in the $|1\rangle$ state as a function of the current in the SG wires. For each value of the current, we measured the position of the atoms along the z-axis as a function of time while being held, and fitted each data set to a parabola. The holding current is adjusted so that the total acceleration is equal to zero, namely, the acceleration due to the holding pulse cancels the acceleration due to gravity and the residual magnetic gradient in the chamber. The measurement is shown in Fig.\,\ref{fig:holding_current_calibration}, and yields a value of $I_{\rm hold} = 23\,\rm\pm 1.5 mA$. As mentioned in the methods section of the main text, while the inaccuracy in satisfying the levitation condition is 6.5\%, it corresponds to only 0.42\% inaccuracy in the phase accumulation, as the deviation from the levitation condition is introduced squared to the phase.

In order to reduce the effect of the ambient gradient in the chamber, and improve the response of the current source, we set the current between the pulses to a small positive value (and not zero), which we call the idle current. The idle current is set to a value of $I_{\rm idle} = 0.47\,\rm mA$, and is on during the entire interferometer sequence.
Including the idle current, we get the following requirement for the kick current to satisfy the closing condition of the interferometer:
\begin{equation}\label{eq:equal_area}
    \left(I_{\rm kick} - I_{\rm idle}\right) 2\tau_{\rm kick} = ({I_{\rm hold} - I_{\rm idle}  }){T_{h}}
\end{equation}

\begin{equation}\label{eq:I_kick_with_idle}
    I_{\rm kick}  = \frac{I_{\rm hold} - I_{\rm idle}  }{2 \tau_{\rm kick}}T_{h} + I_{\rm idle}.
\end{equation}
The idle current does not affect the levitation condition and is also included in the numerical simulations.

\subsection{Magnetic potential energy of the atoms}\label{sec:Magnetic potential}
The potential in the SGI is the magnetic potential and gravity. The magnetic potential is given by
\begin{equation*}
    V_B(\mathbf{r},t) = -\mu\cdot B(\mathbf{r},t).
\end{equation*}
Under the adiabatic approximation, we can assume that the atomic spin projection on the quantization axis, defined by the direction of the magnetic field, is conserved, and adding the second order Zeeman splitting, we get
\begin{equation}\label{eq:Zeeman}
    V_B(\mathbf{r},t) = m_F g_F \mu_B |B(\mathbf{r},t)| + \alpha_{\rm F,m_F}|B(\mathbf{r},t)|^2
\end{equation}
where $g_F \mu_B / h = 0.70\, \rm [MHz/G]$ and $\alpha_{\rm F,m_F}$ is a coefficient that depends on $F$ and $m_F$, given in table\,\ref{table:alpha}.
Using Eq.\,\ref{eq:Zeeman}, the values in table\,\ref{table:alpha} and the measured transition frequency between  $|F=2,m_F=2\rangle$ and $|F=2,m_F=1\rangle$ of $8.799\,\rm MHz$ we can calculate the magnitude of the bias field $|B| = 12.62$\,G. To calculate $a_{SOZ}$, we find the magnetic gradient required to hold the $|F=2,m_F=1\rangle$ state against gravity, and use it to get the value of $a_{SOZ} = 0.10 \rm\,m/s^2$.

\begin{table}[H]
\begin{center}
\begin{tabular}{||c | c c c||}
 \hline
\backslashbox{F}{$m_F$}& 0 & $\pm$1 & $\pm$2 \\
 \hline
 1 &  -287.6 & -215.7 & --- \\
 \hline
 2 &  287.6 & 215.7 & 0 \\
 \hline
\end{tabular}
\caption{Values of $\alpha_{\rm F,m_F} / h$ in units of $\rm Hz/{G}^2$, for calculating the second order Zeeman energy in Eq.\,\ref{eq:Zeeman}.}
\label{table:alpha}
\end{center}
\end{table}

\subsection{Ambient gradient in the chamber}\label{sec:ambient_gradient}
\begin{figure}[H]
\centering
\includegraphics[width=1\textwidth]{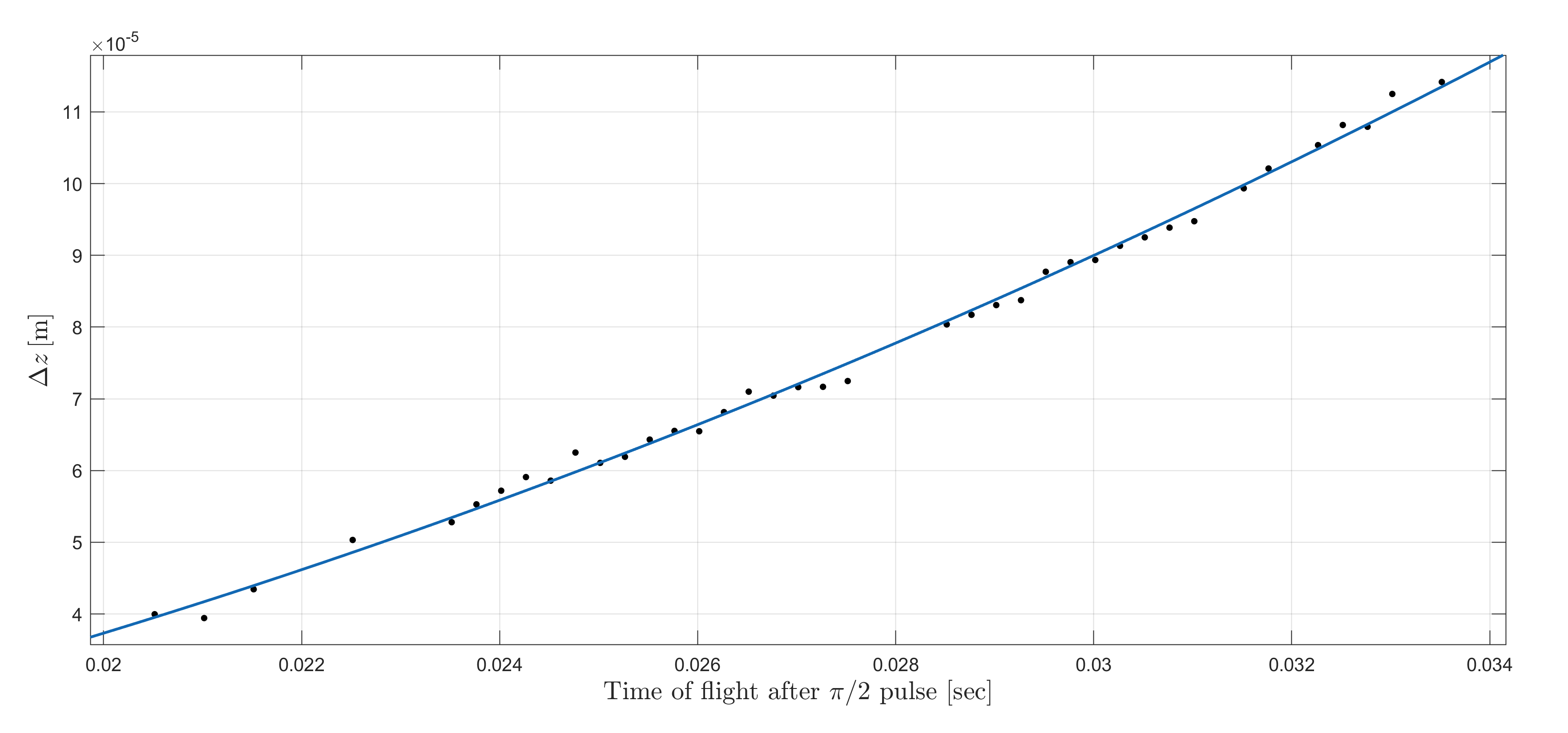}
\caption[]{Spatial SG splitting between the state $|1\rangle$ and $|0\rangle$ due to the ambient gradient, vs the time of flight after the $\pi/2$ pulse. Fitting the data to $0.5a_{ambient}TOF^2$ gives a value of $a_{ambient} = 0.211\pm0.005\,\rm m/s^2$}
\label{fig:Splitting due to ambient gradient vs TOF after pi half pulse.png}
\end{figure}

\begin{figure}[H]
\centering
\includegraphics[width=1\textwidth]{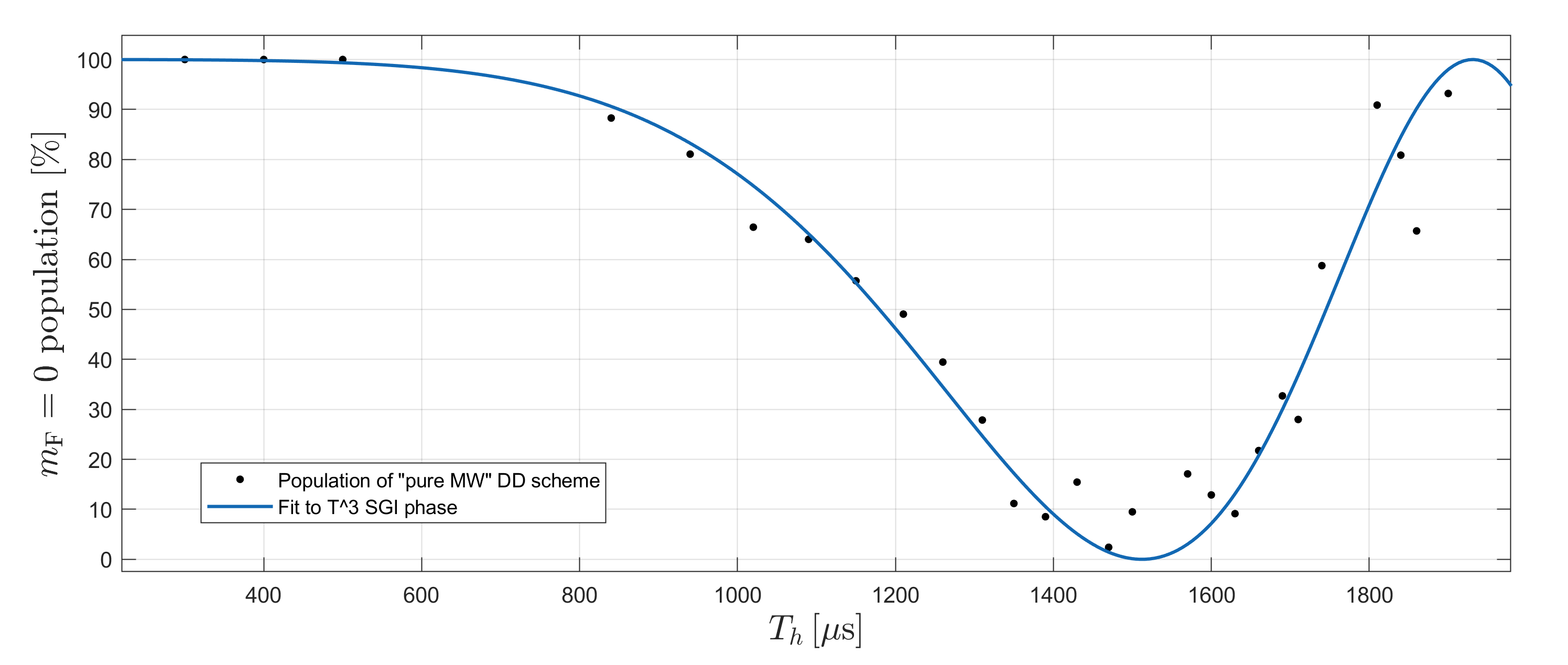}
\caption[]{$T^3$ population oscillations in a pure MW DD scheme. By fitting the data to
$P(\tau_{DD}) = 50+50\cos\left[\frac{m}{\hbar}\tau_{DD}^3\left(a_{ambient}^2 +2ga_{ambient}\right)\right]$, we can estimate the acceleration due to the ambient gradient, and get $a_{ambient} = 0.221\pm0.005\,\rm m/s^2$.}
\label{fig:MW phase and fit to T^3 sgi.png}
\end{figure}

We found a small ambient magnetic gradient in the chamber even when the currents on the atom chip were zero. We measured this ambient magnetic gradient in two methods. The first method is measuring the differential acceleration between the $|1\rangle$ and $|0\rangle$ states in a long time of flight of $20-34 \,\rm ms$, after the $\pi/2$ pulse and observing the spatial splitting of the states. We fit the spatial splitting vs time and find $a_{ambient} = 0.211\pm0.005\,\rm m/s^2$. The data and the fit appear in Fig.\,\ref{fig:Splitting due to ambient gradient vs TOF after pi half pulse.png}. The second method is observing a $T^3$ phase accumulation when measuring the pure MW DD scheme (without magnetic gradients from chip wires) for a duration of $4\tau_{DD} = 4\,\rm ms$, which can, in the presence of the ambient gradient, create a ``self emerging" $T^3$ full-loop SGI. In the original $T^3$ SGI scheme of \cite{Amit2019}, the loop is closed by inverting the orientation of the magnetic gradients. In this scheme, the magnetic gradients remain constant, and the $\pi$ pulses invert the spin of the arms to create a closed interferometer. We calculate the prediction for the phase by plugging the durations $(\tau_{DD},2\tau_{DD},\tau_{DD})$ and accelerations $a_1 = (a_{ambient},0,a_{ambient})+g$, $a_2 = (0,a_{ambient},0)+g$ of the scheme into Eq.\,\ref{eq:action} which gives:
\begin{equation}
   \delta\phi = \frac{m}{\hbar}\tau_{DD}^3\left(a_{ambient}^2 + 2ga_{ambient}\right)\,,
\end{equation}
where $\tau_{DD}=\left(T_h+2  T_d\right) / 2$. Using the prediction for the phase we extract $a_{ambient} = 0.221\pm0.005\,\rm m/s^2$, which corresponds to a magnetic gradient of $0.68 \pm 0.02\,\rm G/cm$. The data and the fit of this measurement appear in Fig.\,\ref{fig:MW phase and fit to T^3 sgi.png}

\subsection{Wave-packet size}

Direct experimental measurement of the wave-packet size is a hard task. To begin with, our optical resolution is only $3\,\rm\mu m$. Second, imaging near the chip distorts the image due to diffraction from the chip. Diffraction also originates from the high optical density of the atom cloud. Finally, the magnetic potential of the DKC, as well as the magnetic fields during the SG kicks, are curved as they emanate from the nearby chip wires. These curvatures create lensing (focusing) effects which result in fast changes of the wave-packet size over time. Given these restrictions, when imaging the wave-packet width we observe a width of $3-4\,\rm\mu m$ on the CCD, which indicates that the actual size is most probably below our optical resolution.

\begin{figure}[H]
\centering
\includegraphics[width=1\textwidth]{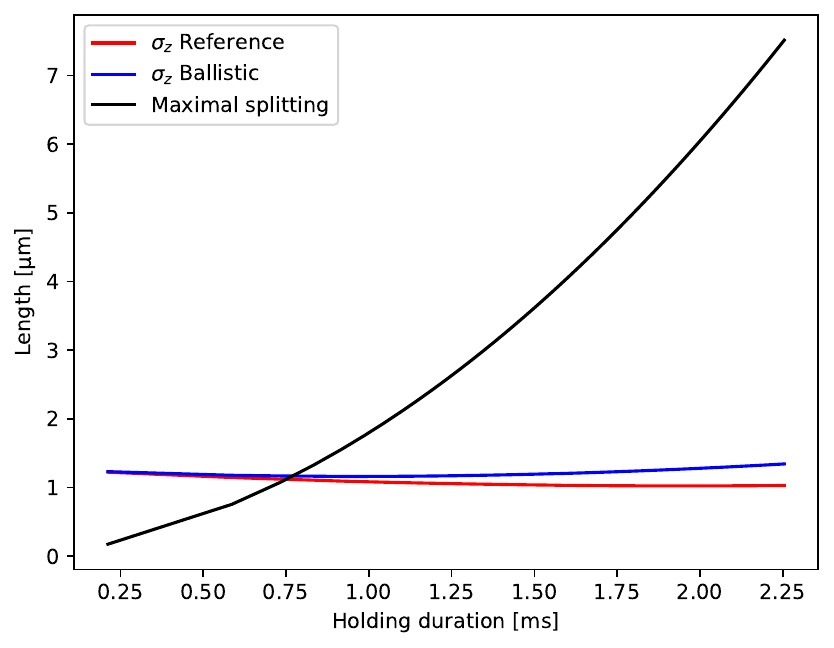}
\caption[]{Numerical calculation of the wave-packet size and the maximal splitting for different interferometer durations. The wave-packet width is defined by the Gaussian width of the WP along the z axis. At the largest holding duration the maximal splitting is $7.5\,\rm\mu m$, and the wave-packet widths are $1\,\rm\mu m$ and $1.3\,\rm\mu m$, for the reference and ballistic WPs respectively.}
\label{fig:WP size}
\end{figure}

A better estimation of wave-packet size is derived using our accurate numerical simulation (Fig.\,\ref{fig:WP size}). The simulation shows that at half way through the interferometer, the wave-packet size is in the range of $1 - 1.3\,\rm\mu m$. The maximum splitting achieved is thus 6.5 times the wave-packet width.

Achieving a larger splitting is feasible but would require significant changes in the experiment, even if we assume that environmental decoherence will not increase for the larger splitting and for the longer duration which is associated with it. The main reason that achieving larger splitting is hard is due to a diminishing overlap and contrast emanating from differential changes of wave-packet size and shape formed by the magnetic potential curvatures discussed above. The larger the required magnetic kicks are, the stronger these effects become. Hence, one would have to create larger gradients (for larger splitting) while suppressing the field curvature. While the curvature goes like $I/r^3$ (r is the distance to the chip and $I$ is the current), the gradient goes like $I/r^2$, so in order to increase the gradient by $2(10)$ while suppressing the curvature by $2(10)$, the distance to the chip would have to be increased by a factor of $4(10^2)$ and the current would have to increase by a factor of $32(10^5)$. For various technological barriers, even the smaller numbers pose a significant challenge.

\subsection{Future experiments with nano-particles}

It should be pointed out that the considerations of this paper refer to the gravitational field as a “passive” background, namely the Earth’s gravitational field, which is not being actively affected by the quantum-mechanical constituents of the atom in the experiment itself. Such influences would be far too tiny to have any effect on the behaviour of the system under consideration.

The distinction between active and passive mass\,\cite{RevModPhys.29.423} goes back to Newton, when he compared the Earth's active mass on the falling apple with the Earth’s passive mass in its motion around the Sun, in its response to the Sun’s active mass.

The D\'iosi-Penrose proposal for the life-time for quantum state-reduction arises from the active effect of bodies in quantum superposition
and represents a vital area of future experimentation. It is argued that when mass plays an active role in the quantum superpositions then the quantum evolution becomes problematic, suggesting a time-scale for quantum state reduction. This could be regarded as a conflict with the equivalence principle as formulated by Marletto and Vedral\,\cite{marletto2020}. In fact, the equivalence principle, as normally considered refers only to the passive effects of mass.

In order to probe the issue of whether active gravitational effects do actually induce the decay of superpositions, one must achieve quantum coherences persisting for at least a second, with mass scales on the order of $10^9$ to $10^{10}$ atomic masses. Presently, the most massive superpositions reached in the laboratory—those involving large molecules—remain at least six to seven orders of magnitude below this threshold\,\cite{RevModPhys.97.015003}. Ongoing experimental efforts, employing nanoparticles, micromechanical mirrors, and nanodiamonds, aim to bridge this gap and reach the necessary regime.




\begin{thebibliography}{}
\bibitem{Galileo-footnote} As an interesting contextual anecdote, one may say that the mathematical description of free fall, as discovered by Galileo, is governed by powers of t: the acceleration is proportional to $t^0$, the velocity to $t^1$, and the position to $t^2$. In this work we extend this framework and demonstrate experimentally that the quantum phase of a freely falling mass follows a $t^3$ power law.
\bibitem{frames-footnote} Let us note that the ballistic frame [as defined by Fig.\,1(A)] and the standard freely-falling frame (monotonically increasing velocity towards Earth) are the same up to the constant velocity $v_0$. As physics is invariant to such a constant velocity shift, our ballistic frame may be regarded as a freely-falling frame.
\bibitem{observable-footnote} If one computes the relative phase between $\phi_{\text{ballistic}}$ and $\phi_{\text{reference}}$  in each of the frames, this relative phase comes with a different sign depending on the frame in which the calculation was made. This sign has no observable effect in our setup as the observable interference fringes are the result of a cosine function which is not sensitive to the sign.
\end{thebibliography}
\end{document}